\documentclass[12pt, a4paper]{article}
\pdfoutput=1

\usepackage[height=22cm,width=17cm, centering]{geometry}
\usepackage{setspace}

\usepackage[utf8]{inputenc} 

\usepackage{amsmath}
\usepackage{amssymb, mathrsfs}
\usepackage{graphicx}
\usepackage{subcaption}
\usepackage{color,comment, slashed, bm, textcomp}
\usepackage[sort&compress, numbers, merge]{natbib}

\definecolor{purple}{rgb}{0.65 ,0, 0.75}
\definecolor{red}{rgb}{0.8, 0., 0.3}
\definecolor{orange}{rgb}{0.8, 0.3, 0}
\usepackage{hyperref}
\hypersetup{colorlinks=true, linkcolor=red, citecolor=orange, urlcolor=purple}

\def \Mpl {M_{\text{P}}}

\allowdisplaybreaks

\begin{document}
\begin{titlepage}
\hfill {\small CTPU-PTC-21-38}\\
\centering
\vspace{3 cm} \\
{\LARGE Constrained Superfields in Dynamical Background} \\
\vspace{1 cm}
\renewcommand{\thefootnote}{\fnsymbol{footnote}}
{ Shuntaro Aoki$^1$\footnote{\href{mailto:shuntaro@cau.ac.kr}{\texttt{shuntaro@cau.ac.kr}}} and Takahiro Terada$^2$\footnote{\href{mailto:takahiro@ibs.re.kr}{\texttt{takahiro@ibs.re.kr}}}} \\
\setcounter{footnote}{0} 
\vspace{0.6 cm}
$^1$  Department of Physics, Chung-Ang University, Seoul 06974, Republic of Korea \\
$^2$  Center for Theoretical Physics of the Universe, \\ Institute for Basic Science, Daejeon 34126, Republic of Korea
\vspace{1cm}
\begin{abstract}
We study the nonlinear realization of supersymmetry in a dynamical/cosmological background in which derivative terms like kinetic terms are finite.  Starting from linearly realized theories, we integrate out heavy modes without neglecting derivative terms to obtain algebraic constraints on superfields.  Thanks to the supersymmetry breaking contribution by the kinetic energy, the validity of constrained superfields can be extended to cosmological regimes and phenomena such as reheating after inflation, kinetic-energy domination, and the kinetic and standard misalignment of axion.
\end{abstract}

\end{titlepage} 

\section{Introduction \label{sec:intro}}

Physicists often approximate things by the most symmetric configurations as a starting point. The most symmetric spacetime is homogeneous and isotropic; it has Poincare invariance and may even be supersymmetric.  
Nevertheless, interesting phenomena are often associated with dynamics and hence derivatives of fields. Dynamical features such as oscillations of scalar fields appear, e.g., in cosmology. Note that nonzero derivative terms like the kinetic term breaks supersymmetry spontaneously.  The importance of such terms in supersymmetry and supergravity has been emphasized in the cosmological context~\cite{Kallosh:2000ve}.  We extend this philosophy to nonlinear realization of supersymmetry~\cite{Volkov:1972jx, Volkov:1973ix, Rocek:1978nb, Ivanov:1978mx, Lindstrom:1979kq, Casalbuoni:1988xh, Komargodski:2009rz, Kuzenko:2010ef}.

Well below the scale of supersymmetry breaking, supersymmetry is realized nonlinearly. Similar to the fact that the restriction of the linear sigma model to the non-linear sigma model is given by a manifestly symmetric constraint equation, the corresponding constraint in supersymmetry is invariant under supersymmetry transformation~\cite{Casalbuoni:1988xh}. A famous example is the nilpotency condition $\bm{S}^2 = 0$ for a chiral superfield $\bm{S}$~\cite{Rocek:1978nb, Casalbuoni:1988xh, Komargodski:2009rz}.  With such an algebraic constraint, the description of the low-energy physics in terms of superfields is possible, and resultant interactions are highly restricted (or predictive) due to the nilpotency.  Thus, constrained superfields are convenient tools to describe spontaneously broken supersymmetry. 

Constrained superfields (see Ref.~\cite{Farakos:2017bxs} for a review) were used in the context of de Sitter~\cite{Bergshoeff:2015tra, Hasegawa:2015bza, Kallosh:2015sea}, moduli stabilization~\cite{Aparicio:2015psl},  inflation~\cite{Antoniadis:2014oya, Ferrara:2014kva, Kallosh:2014via, Kallosh:2014hxa, DallAgata:2014qsj, Kahn:2015mla, Ferrara:2015tyn, Carrasco:2015iij, Dudas:2016eej, Argurio:2017joe, Dalianis:2017okk}, and late-time universe~\cite{Burgess:2021juk}.  They were studied also in the context of extended supersymmetry~\cite{Kuzenko:2011ya, Kuzenko:2015rfx, Cribiori:2016hdz, Dudas:2017sbi, Antoniadis:2017jsk, Kuzenko:2017zla, Aldabergenov:2021rxz} and in superstring theory~\cite{Kallosh:2015nia, Bandos:2015xnf,  Garcia-Etxebarria:2015lif, Vercnocke:2016fbt, Kallosh:2016aep, Aalsma:2017ulu, Kuzenko:2017gsc, Kallosh:2018nrk, Cribiori:2019hod, Cribiori:2020bgt}.  By using constrained superfields, heavy irrelevant modes are automatically decoupled.  This means, e.g., in the context of inflation, one does not need to worry about light isocurvature modes otherwise present in the model.  This procedure is actually nontrivial.  From the purely low-energy perspective, imposing a constraint by hand is perfectly fine.   However, if we assume that there is a description with linearly realized supersymmetry, the constraints should be consistent with the UV description.\footnote{
One can implement the constraint in the linearly realized theory using Lagrange multipliers~\cite{Ferrara:2016een}, but we would categorize it as one of the low-energy approaches.
}  
In fact, Refs.~\cite{Dudas:2011kt, Antoniadis:2011xi, Antoniadis:2012ck} (see also Ref.~\cite{Ghilencea:2015aph} for a review and Ref.~\cite{Cribiori:2017ngp} for a counter perspective) showed that the nilpotent superfield is not always obtained as an automatic consequence of supersymmetry breaking, but it depends on the couplings in the K\"ahler potential in the UV theory.  
Therefore, it is important to check the UV consistency of constrained superfields as well as their applicable range and to clarify which types of constraints appear in low energy from the UV perspective.

In this paper, we derive known and new constraints on superfields from some UV setups following the approach of Refs.~\cite{Dudas:2011kt, Antoniadis:2011xi, Antoniadis:2012ck}, {\emph{without}} neglecting derivative terms. This is in contrast to the previous works where such derivative terms are neglected by assuming that they do not much contribute to supersymmetry breaking.\footnote{The constraints with derivative terms are studied in the context of partial $\mathcal{N}=2\rightarrow 1$ supersymmetry breaking, see e.g., Refs.~\cite{Bagger:1996wp,Bagger:1997pi,Rocek:1997hi,Gonzalez-Rey:1998vtf}. } The superfield constraints obtained in our work are widely applicable to cosmological scenarios with dynamical backgrounds due to non-trivial kinetic terms of a scalar field. They  include not only supersymmetry breaking at vacuum and de Sitter inflationary phase but also (p)reheating after inflation~\cite{Dolgov:1989us, Traschen:1990sw, Kofman:1994rk, Shtanov:1994ce, Kofman:1997yn, Greene:2000ew, Amin:2014eta}, kinetic-energy dominated universe (kination)~\cite{Spokoiny:1993kt, Joyce:1996cp, Ferreira:1997hj}, the curvaton mechanism~\cite{Linde:1996gt, Enqvist:2001zp, Lyth:2001nq, Moroi:2001ct}, the kinetic~\cite{Co:2019jts} and standard~\cite{Preskill:1982cy, Abbott:1982af, Dine:1982ah} axion misalignment, cosmological relaxation~\cite{Abbott:1984qf, Graham:2015cka}, etc., which opens up new possibilities for model building based on constrained superfields.     

The structure of the paper is as follows.  In Section~\ref{sec:single}, we begin with a brief review of the derivation of the nilpotent superfield from a UV model, followed by its generalization to the shift-symmetric case. We explicitly show that the cubic shift-symmetric constraint~\cite{Komargodski:2009rz, Aldabergenov:2021obf} emerges in the low-energy theory. We also discuss its relation to inflation mechanisms with a single chiral superfield.  In Section~\ref{sec:multi}, we extend the discussion into the double superfield case. We focus on the setup well-motivated by inflation-model building.  We find novel quintic constraints after integrating out the heavy modes. Implications on cosmological particle production are discussed. These two sections illustrate our main ideas. Generalizations into supersymmetric gauge theories and supergravity are discussed in Section~\ref{sec:generalization}.  Section~\ref{sec:conclusion} is devoted to our summary and conclusion. Some technical details and side remarks are included in three appendices.

We consider four-dimensional $\mathcal{N}=1$ supersymmetry and basically follow the conventions of Ref.~\cite{Wess:1992cp}. A superfield is denoted by a bold-face letter.  We take the natural unit $c = \hbar = 1$ and occasionally use $M_\text{P} \equiv 1/\sqrt{8\pi G} = 1$.

\section{Single superfield \label{sec:single}}
In this section, we consider some UV models with a single superfield and derive supersymmetric constraints keeping derivative terms as they are. Starting from a review of the standard nilpotent chiral superfield, we discuss a model that leads to the cubic shift-symmetric constraint and its application to inflation.

\subsection{Review of nilpotent chiral superfield \label{sec:nilpotent}}
When supersymmetry is broken, some fields get heavy and decouple from the low-energy dynamics.  Well below the mass scales of the decoupled fields, it is realized nonlinearly. 
For our purposes, it is useful to consider the simplest example with a single chiral superfield $\bm{S} = S + \sqrt{2} \theta \chi + \theta \theta F$ that breaks supersymmetry spontaneously.  Let us consider a UV theory in which supersymmetry is linearly realized with the following K\"ahler potential\footnote{
 In the normal coordinates of the K\"ahler manifold, $K_{SS \bar{S}} = K_{SSS\bar{S}} = 0$ and $K_{S\bar{S}} = 1$~\cite{Antoniadis:2011xi}, where subscripts denote differentiation.
} and superpotential:
\begin{align}
    K (\bm{S}, \bar{\bm{S}}) = & \bar{\bm{S}}\bm{S} - \frac{1}{4 \Lambda^2} \left( \bar{\bm{S}}\bm{S} \right)^2, \label{K_S}\\
    W (\bm{S}) =& f \bm{S}. \label{W_S}
\end{align}
Here, the superpotential is linear in $\bm{S}$ to break supersymmetry spontaneously.  The role of the quartic term in $K$ -- the K\"ahler curvature -- is to make the scalar component $S$ so heavy that it decouples. 
The Lagrangian density in terms of the component fields is 
\begin{align}
    \mathcal{L} =& K_{S \bar{S}}  \left( - \partial^\mu \bar{S} \partial_\mu S
 - \frac{i}{2} \left(\chi \sigma^\mu D_\mu \bar{\chi} - D_\mu \chi \sigma^\mu \bar{\chi} \right) + F\bar{F}   \right) \nonumber \\
& + \left( \left(W_S - \frac{1}{2} K_{S \bar{S}\bar{S}} \bar{\chi}\bar{\chi} \right) F - \frac{1}{2} W_{SS} \chi \chi \right) + \text{H.c.} + \frac{1}{4} K_{SS\bar{S}\bar{S}} \chi\chi \bar{\chi}\bar{\chi},  \label{L_single}
\end{align}
where H.c.~denotes the Hermitian conjugate. 
The covariant derivative is defined as $D_\mu \chi^i = \partial_\mu \chi^i + \Gamma^i_{jk} \partial_\mu \Phi^j \chi^k$ for a general set of chiral superfields $\bm{\Phi}^i$ with $\Gamma^i_{jk} = K^{\bar{\ell}i} K_{j k \bar{\ell}}$ being the K\"ahler connection and $K^{\bar{j}i}$ the inverse of the K\"ahler metric $K_{i\bar{j}}$.

Now, let us take a formal limit $\Lambda \to 0$~\cite{Casalbuoni:1988xh} so that the mass of the complex scalar $S$ diverges.\footnote{
  Of course, it is not guaranteed to be able to take such a limit in a realistic model since $\Lambda$ may be associated with the mass scale of the underlying UV models.  Nevertheless, analyses in Ref.~\cite{Dudas:2016eej} show that there exists a narrow range of parameters in the underlying model in which this procedure is valid and applicable to inflation.
}
The equation of motion for the heavy field $S$ is dominated by the part proportional to the K\"ahler curvature ($\sim 1/\Lambda^2$),
\begin{align}
S \bar{F}F =& \frac{1}{2} \bar{F} \chi \chi  - i \partial_\mu \left(  \chi \sigma^\mu \bar{\chi} S \right)  - \frac{1}{2} \bar{S} \Box (S^2) +iS\chi\sigma^\mu \partial_{\mu}\bar{\chi} .
\end{align}
In the following, we focus on a solution for $S$ which is given by fermion bilinears and higher-order terms in fermions, if any. Thus, we neglect the purely bosonic contribution to the solution.\footnote{The same comment applies to the remaining sections.} 
Then, the equation for $S$ is solved as~\cite{Casalbuoni:1988xh, Komargodski:2009rz} 
\begin{align}
    S = \frac{\chi\chi}{2 F}, \label{nilpotent_solution}
\end{align}
provided that $F \neq 0$. 
Note that we have not dropped the derivative terms by hand but they vanish identically because of the fermionic feature ($\mathcal{O}(\chi^3) = 0$). 
This implies $S^2= 0$, and its supersymmetry transformation also vanishes with this constraint. In general, when the lowest component of a superfield and its supersymmetry transform vanish, the whole superfield vanishes.\footnote{
In the case of the Wess-Zumino gauge of a real superfield, the supersymmetry transformation must be accompanied by the compensating gauge transformation to keep the Wess-Zumino gauge.  Because of this modification of the transformation law, our statement here does not directly apply to the case of the Wess-Zumino gauge.
}  Therefore, supersymmetry tells us that
\begin{align}
    \bm{S}^2 = 0, \label{nilpotency}
\end{align}
i.e., the superfield $\bm{S}$ is nilpotent~\cite{Casalbuoni:1988xh, Komargodski:2009rz}. Note that the $F$-term of this constraint gives eq.~\eqref{nilpotent_solution}, while the other components do not yield new constraints. 
The constraint gives rise to the low-energy theory with spontaneously broken supersymmetry, but the constraint equation itself is supersymmetric.  This is analogous to the constraint in the non-linear sigma model~\cite{Casalbuoni:1988xh}.

We can see this also in the superfield language. The equation of motion in superspace is $ \bar{W}_{\bar{\bm{S}}} - \frac{1}{4} \mathscr{D}^2 K_{\bar{\bm{S}}} = 0$, where $\mathscr{D}$ denotes the superderivative. The part enhanced by the K\"ahler curvature in the limit $\Lambda \to 0$ is \begin{align}
    \bar{\bm{S}} \mathscr{D}^2 \bm{S}^2 = 0.
\end{align}
Assuming the nowhere vanishing $F$-term, $\mathscr{D}^2 \bm{S} \neq 0$, we can rewrite it as follows:
\begin{align}
    \bm{S} = &  - \frac{\mathscr{D}\bm{S} \mathscr{D}\bm{S}\bar{\mathscr{D}}^2 \bar{\bm{S}}+ 4 \bar{\mathscr{D}} \bar{\bm{S}} \bar{\mathscr{D}} \mathscr{D} \bm{S} \mathscr{D} \bm{S} +\bar{\bm{S}} \left( 2 \bar{\mathscr{D}} \mathscr{D} \bm{S} \bar{\mathscr{D}}\mathscr{D} \bm{S} + \bm{S} \bar{\mathscr{D}}^2 \mathscr{D}^2 \bm{S} \right)}{\mathscr{D}^2 \bm{S} \bar{\mathscr{D}}^2 \bar{\bm{S}} + 2 \bar{\mathscr{D}}\bar{\bm{S}} \bar{\mathscr{D}}\mathscr{D}^2 \bm{S}}.
\end{align}
Using the anticommutation relation of the superderivatives and solving this equation recursively, one can see that eq.~\eqref{nilpotency} follows. 

The nilpotency condition~\eqref{nilpotency} is often imposed by hand (or equivalently by the   Lagrange multiplier) in inflationary applications~\cite{Antoniadis:2014oya, Ferrara:2014kva, Kallosh:2014via, Kallosh:2014hxa, DallAgata:2014qsj, Kahn:2015mla, Ferrara:2015tyn, Carrasco:2015iij, Dalianis:2017okk}.  Here, we have reviewed its derivation from a UV model, and the derivation implies possible correction terms suppressed by the mass of the heavy particle $S$. 

In this model, the only remaining degree of freedom in the low-energy is the Nambu-Goldstone fermion (goldstino) $\chi$.  Its superpartner, the sgoldstino $S$ has been decoupled.  The low-energy action is equivalent to the Akulov-Volkov action~\cite{Volkov:1972jx, Volkov:1973ix} via a field redefinition~\cite{Kuzenko:2010ef}.

It is interesting to comment on the fact that the denominator in eq.~\eqref{nilpotent_solution} is the $F$-term supersymmetry breaking scale $F = - \bar{f} + \cdots$, where dots represent fermionic terms.  The expression~\eqref{nilpotent_solution} is valid only when $F$ is nonvanishing. In inflationary applications, the nilpotent superfield $\bm{S}$ is often used with a modification $f \to f(\bm{\Phi})$ where $\bm{\Phi}$ is the inflaton superfield.  As pointed out in Ref.~\cite{DallAgata:2014qsj}, the description becomes invalid if $f(\Phi)$ vanishes (and it typically does after inflation). Ref.~\cite{DallAgata:2014qsj} introduced a specific superpotential to solve the problem.  Instead, we consider an alternative possibility in which the $F$-term in the denominator in eq.~\eqref{nilpotent_solution} is augmented by the kinetic energy of inflaton.  We will discuss this interesting possibility and obtain novel constrained superfields in Section~\ref{sec:multi}.  Before discussing such a multi-superfield scenario, we will now discuss the minimal example with the essentially same mechanism.

\subsection{Shift symmetry and cubic constraint \label{sec:cubic}}

For the kinetic energy of a scalar field to play a major role in the low-energy theory, we need a scalar field that is not decoupled.  Suppose that the scalar field is protected by an approximate global shift symmetry.  We rename the supersymmetry breaking field $\bm{\Phi}$ and consider its shift symmetric K\"ahler potential $K(\bm{\Phi}, \bar{\bm{\Phi}}) = K ( \bm{\Phi} + \bar{\bm{\Phi}})$ invariant under the imaginary shift of $\bm{\Phi}$. The shift symmetry may originate, e.g., from a U(1) symmetry or the dilatation symmetry, and our discussion can be relevant for an axion or a dilaton, respectively, or any (axionic or dilatonic) light field such as an inflaton.

The component Lagrangian is obtained simply by relabeling $S \to \Phi (\equiv  \frac{1}{\sqrt{2}}(\phi + i \varphi))$ in eq.~\eqref{L_single}. We take the shift-symmetric version of eq.~\eqref{K_S},
\begin{align}
    K (\bm{\Phi},\bar{\bm{\Phi}}) =& \frac{1}{2}(\bm{\Phi}+\bar{\bm{\Phi}})^2 - \frac{1}{4! \Lambda^2} (\bm{\Phi} + \bar{\bm{\Phi}})^4. \label{K_shift-symmetric}
\end{align}
An analysis for the generic $K$ is given in Appendix~\ref{sec:general_EOM} including the effects of shift-symmetry breaking.
The quartic term gives a mass to the real part $\phi$, whereas the imaginary part $\varphi$ remains light.

In the following, we integrate out the heavy mode $\phi$ and show that $\bm{\Phi}$ satisfies a constraint $(\bm{\Phi} + \bar{\bm{\Phi}})^3 = 0$~\cite{Aldabergenov:2021obf} in the limit of the large K\"ahler curvature.
The terms in equations of motion enhanced by the K\"ahler curvature satisfy
\begin{align}
    & \sqrt{2}\phi \left( - \frac{1}{2} \partial^\mu \varphi \partial_\mu \varphi + \bar{F}F - \frac{i}{2}\left(\chi \sigma^\mu \partial_\mu \bar{\chi} - \partial_\mu \chi \sigma^\mu \bar{\chi} \right) \right) \nonumber \\
    =& \frac{\bar{F}}{2}\chi \chi + \frac{F}{2} \bar{\chi}\bar{\chi} + \frac{1}{\sqrt{2}}\chi \sigma^\mu \bar{\chi} \partial_\mu \varphi - \frac{1}{\sqrt{2}} \phi^2 \Box \phi - \frac{1}{\sqrt{2}} \phi \partial^\mu \phi \partial_\mu \phi. \label{EoM_phi-only}
\end{align} 
In the same way as before, this can be solved recursively starting at the quadratic fermion terms.  In fact, the leading-order terms with respect to the fermion number are
\begin{align}
   \phi = \frac{1}{\sqrt{2}\rho} \left( \frac{\bar{F}}{2} \chi \chi + \frac{F}{2} \bar{\chi}\bar{\chi} + \frac{1}{\sqrt{2}} \chi\sigma^\mu \bar{\chi} \partial_\mu \varphi \right) + \cdots, \label{phi-only_sol_leading}
\end{align}
where $\rho \equiv  - \frac{1}{2} \partial^\mu \varphi \partial_\mu \varphi + \bar{F}F $ is the background ``energy density''\footnote{ \label{fn:energy_density}
We are assuming that the spatial inhomogeneity is negligible. Too large gradient energy leads to a singularity $(\rho =0)$~\cite{Aldabergenov:2021obf}.
} or the order parameter of the supersymmetry breaking.  This tells us that $\phi^3 = 0$, which is again compatible with supersymmetry.  That is, its supersymmetry transformation vanishes under this constraint. From supersymmetry, the following superfield constraint is obtained
\begin{align}
    (\bm{\Phi} + \bar{\bm{\Phi}})^3 = 0. \label{cubic_constraint}
\end{align}
In fact, $\phi^3 = 0$ implies $\phi^2 \partial_\mu \phi = 0$ and $\phi^2 \Box \phi + 2 \phi \partial^\mu \phi \partial_\mu \phi = 0$, etc.  Using the latter, we see that eq.~\eqref{EoM_phi-only} reproduces eqs.~(15--18) of Ref.~\cite{Aldabergenov:2021obf}.  The complete solution of eq.~\eqref{EoM_phi-only} is given in Ref.~\cite{Aldabergenov:2021obf} and we have confirmed it. Thus, we have \emph{derived} from UV the constraint $(\bm{\Phi}+\bar{\bm{\Phi}})^3 = 0$ rather than \emph{imposed} it at IR by hand.

We comment on other cubic or higher constraints in the literature. The nilpotency constraint~\eqref{nilpotency} becomes cubic or higher, e.g., $\bm{S}_1^3 = \bm{S}_1^2 \bm{S}_2 = \bm{S}_1 \bm{S}_2^2 = \bm{S}_2^3 = 0$  when multiple superfields break supersymmetry and they couple to each otehr in the K\"ahler curvature term~\cite{Dudas:2011kt, Antoniadis:2011xi, Antoniadis:2012ck, Ghilencea:2015aph}, which was studied without shift symmetry.  The constraint with the same form as eq.~\eqref{cubic_constraint} arises for orthogonal nilpotent superfields $\bm{X}$ and $\bm{A}$~\cite{Komargodski:2009rz, Kahn:2015mla, Ferrara:2015tyn, Carrasco:2015iij, DallAgata:2015zxp}.  They satisfy $\bm{X}^2 = \bm{X} (\bm{A} + \bar{\bm{A}}) = 0$, from which $(\bm{A} + \bar{\bm{A}})^3 = 0$ follows. See Refs.~\cite{Komargodski:2009rz, Aldabergenov:2021obf} for more about this relation. We have another comment on the orthogonal nilpotent superfields at the end of this section.  A cubic constraint was discussed also for a deformed real linear superfield~\cite{Kuzenko:2017oni} and in $\mathcal{N}=2$ supersymmetry~\cite{Kuzenko:2017zla, Dudas:2017sbi, Kuzenko:2017gsc}.

An analogous discussion is possible in superspace. The equation of motion in the large K\"ahler curvature limit is 
\begin{align}
    (\mathscr{D}^2 + \bar{\mathscr{D}}^2) (\bm{\Phi} + \bar{\bm{\Phi}} ) ^3 = 0.
\end{align}
Applying $\bar{\mathscr{D}}^2$, we can express $(\bm{\Phi} + \bar{\bm{\Phi}})$ in terms of superderivatives of $\bm{\Phi}$ and its conjugate. A manifestly real expression is 
\begin{align}
   (\bm{\Phi} + \bar{\bm{\Phi}}) = & - \frac{4 \left( \bar{\mathscr{D}}\bar{\bm{\Phi}}\bar{\mathscr{D}}\bar{\bm{\Phi}} \mathscr{D}^2 \bm{\Phi}+ \text{H.c.} \right)+ 8 \bar{\mathscr{D}}\bar{\bm{\Phi}} \mathscr{D}\bm{\Phi} \left(\bar{\mathscr{D}}\mathscr{D}\bm{\Phi} +\text{H.c.} \right)  + (\bm{\Phi} + \bar{\bm{\Phi}})^2 \left( \bar{\mathscr{D}}^2 \mathscr{D}^2 \bm{\Phi} + \text{H.c.} \right)  }{4\mathscr{D}^2\bm{\Phi} \bar{\mathscr{D}}^2 \bar{\bm{\Phi}} +4\left( (\bar{\mathscr{D}}\mathscr{D}\bm{\Phi})^2 + \text{H.c.}  \right) + 4 \left( \bar{\mathscr{D}}\bar{\bm{\Phi}}\bar{\mathscr{D}}\mathscr{D}^2 \bm{\Phi} + \text{H.c.} \right) },
\end{align}
where we assumed that the denominator never vanishes, which is a generalization of the assumption that the $F$-term supersymmetry breaking is nonzero. 
Recursively solving it, one reproduces eq.~\eqref{cubic_constraint}.

Let us compare $\bm{\Phi}$ with the nilpotent superfield $\bm{S}$.  Similar to the previous case, the constraint is obtained in the limit of the large K\"ahler curvature or equivalently the large mass of $\phi$ compared to other mass scales in the theory.  For a finite mass, there are corrections suppressed by the mass of $\phi$ similarly to the case of Refs.~\cite{Dudas:2011kt, Antoniadis:2011xi, Antoniadis:2012ck, Ghilencea:2015aph}.

However, there is an interesting difference: the denominator in eq.~\eqref{phi-only_sol_leading} has the kinetic energy of the light scalar $\varphi$ as well as the usual term $|F|^2$.  After elimination of the auxiliary field by its equation of motion, it becomes $|F|^2 = V(\varphi) + \dots$, where dots represent the terms depending on the goldstino $\chi$, which contribute to higher-order fermion terms of $\phi$.  The sum of the kinetic energy and the potential energy, namely the total background energy, generally behaves more regularly than each of them. It is an adiabatic invariant in the sense that the time derivative is of order the Hubble parameter $H$ times the original quantity $\dot{\rho}/\rho = \mathcal{O}(H)$~\cite{Ema:2015eqa}.

We can take some limits in the solution, \eqref{phi-only_sol_leading}.  First, in the IR limit with vanishing derivatives (as in Refs.~\cite{Dudas:2011kt, Antoniadis:2011xi, Antoniadis:2012ck, Ghilencea:2015aph}), we obtain 
\begin{align}
    \sqrt{2}\phi = \frac{\chi\chi}{2 F} + \frac{\bar{\chi}\bar{\chi}}{2 \bar{F}} + \cdots. \label{phi_sol_F-limit}
\end{align}
This is simply the real part of the well-known nilpotent solution~\eqref{nilpotent_solution}, but we still have $\phi^3 = 0$ with $\phi^2 \neq 0$. 

Second, we take the opposite limit where the $F$-term is negligible.  In this case, supersymmetry is dominantly broken by the kinetic term of $\varphi$, and we have
\begin{align}
    \phi = \frac{\chi \sigma^\nu \bar{\chi}\partial_\nu \varphi}{-\partial^\mu \varphi \partial_\mu \varphi} + \cdots. \label{phi_sol_K-limit}
\end{align}
We cannot take the limit $\rho = 0$ while maintaining the constraint since it would restore supersymmetry and $\phi$ becomes light.

\subsection{Inflationary application and different parametrizations}
Constrained superfields can be useful tools to describe inflation.  First of all, the inflationary energy density necessarily breaks supersymmetry.  Light scalar fields develop quantum fluctuations during inflation, so it is often required to stabilize such light fields to satisfy the bound on isocurvature perturbations~\cite{Planck:2018jri}.  After integrating out the stabilized modes, the remaining degrees of freedom can be conveniently described by constrained superfields. For our shift-symmetric single-superfield example, $\varphi$ can be used as the inflaton.  The scalar superpartner of inflaton (sinflaton) $\phi$ is decoupled and replaced by terms dependent on inflatino/goldstino $\chi$.  

Deep inside the slow-roll inflationary regime, it is a good approximation to neglect derivative terms (the slow-roll approximation) reproducing eq.~\eqref{phi_sol_F-limit}.  After the end of inflation, the inflaton typically oscillates around its minimum, so both its kinetic energy and potential energy are relevant [see eq.~\eqref{phi-only_sol_leading}]. Alternatively, the kinetic energy of the inflaton can dominate to realize the kination era if the potential does not have a minimum as for the runaway-type potential.  In this case, eq.~\eqref{phi_sol_K-limit} becomes relevant. These stories are not restricted to inflation and can apply to other scalar fields like moduli, dilaton, axion, curvaton, Higgs field, quintessence field, etc.

In supergravity, realizing large-field inflation is nontrivial because of the exponential factor and the negative semidefinite term in the scalar potential.  There are several mechanisms to solve these difficulties.  In the following, we discuss which class of models allows the low-energy description by the constrained superfield obeying $(\bm{\Phi} + \bar{\bm{\Phi}})^3 = 0$.

Eq.~\eqref{K_shift-symmetric} is actually nothing but the K\"ahler potential of the mechanism developed in Refs.~\cite{Izawa:2007qa, Ketov:2014qha,Ketov:2014hya,Ketov:2015tpa, Ketov:2016gej} provided that there is an exponential factor $e^{c \Phi}$ with $c \gtrsim \sqrt{3}$ in the superpotential (or equivalently the linear term $c (\Phi + \bar{\Phi})$ in the K\"ahler potential). The K\"ahler curvature term is an integrated component of this class of models and it ensures a positive inflationary potential and stabilize the sinflaton simultaneously.  It can embed various inflation models in supergravity. Inflation using this mechanism can be described by the constrained superfield obying $(\bm{\Phi} + \bar{\bm{\Phi}})^3 = 0$ if $\Lambda$ is sufficiently smaller than the Planck scale. 

If $\bm{\Phi}$ keeps supersymmetry breaking by its $F$-term after inflation, the inflaton is identified as sgoldstino (sgoldstino inflation)~\cite{Achucarro:2012hg, AlvarezGaume:2010rt, AlvarezGaume:2011xv}. The constrained superfield obeying the cubic constraint~\eqref{cubic_constraint} and its UV setup~\eqref{K_shift-symmetric}~\cite{Izawa:2007qa, Ketov:2014qha,Ketov:2014hya,Ketov:2015tpa, Ketov:2016gej} are an explicit realization of sgoldstino inflation. However, there is a severe gravitino problem when the inflaton is identified as sgoldstino in vacuum (see Ref.~\cite{Terada:2021rtp} and references therein).

There are other inflationary models and mechanisms in supergravity with a single superfield~\cite{Goncharov:1983mw, Kumekawa:1994gx, Linde:2014hfa, Roest:2015qya, Linde:2015uga, Scalisi:2015qga, Ferrara:2016vzg, Antoniadis:2017gjr, Antoniadis:2019dpm}.  Without the K\"ahler curvature term, the sinflaton mass is not hierarchically larger than the Hubble scale, so the inflation in such theories cannot be described by the constrained superfield satisfying $(\bm{\Phi} + \bar{\bm{\Phi}})^3 = 0$.  To describe the dynamics after inflation, one has to use the linear description of supersymmetry, which includes the sinflaton $\phi$.

One can modify the K\"ahler potential so that the K\"ahler curvature becomes sizable.  For example, the hyperbolic geometry of the K\"ahler manifold, relevant for the no-scale supergravity~\cite{Cremmer:1983bf, Ellis:1983sf, Lahanas:1986uc} and the $\alpha$-attractor models of inflation~\cite{Kallosh:2013yoa, Galante:2014ifa, Carrasco:2015pla}, can be modified as follows to stabilize the sinflaton~\cite{Carrasco:2015uma} 
\begin{align}
     K = \frac{-3 \alpha}{1 + 2 c_2} \log \left(  \frac{\bm{T}+\bar{\bm{T}}}{2 \sqrt{\bar{\bm{T}}\bm{T}}} \left [ 1 + c_2 \left( \frac{\bm{T}-\bar{\bm{T}}}{\bm{T}+\bar{\bm{T}}} \right)^2 + c_4 \left( \frac{\bm{T}-\bar{\bm{T}}}{\bm{T}+\bar{\bm{T}}} \right)^4  \right ]  \right), \label{modified_hyperbolic_geometry}
\end{align}
where $\alpha$, $c_2$, and $c_4$ are real dimensionless parameters. 
This is scale invariant under $\bm{T} \to c \bm{T}$ and $\bar{\bm{T}} \to c \bar{\bm{T}}$ ($c$: const.). It is possible to check that this model is also in the same category as we studied above by field redefinition $\bm{T} = \exp (2 i \bm{\Phi} / \sqrt{3\alpha})$:
 \begin{align}
 K =&  - \frac{3 \alpha}{2(1 + 2 c_2)}  \left [  - \log 2 +    \log \left( 1 + \cos \frac{2 (\bm{\Phi} + \bar{\bm{\Phi}})}{\sqrt{3\alpha}} \right) \right. \nonumber \\
 & \qquad \qquad  \left. + 2 \log  \left( 1 + c_2 \left( \tan \frac{2 (\bm{\Phi} + \bar{\bm{\Phi}})}{\sqrt{3\alpha}}  \right)^2 + c_4 \left( \tan \frac{2 (\bm{\Phi} + \bar{\bm{\Phi}})}{\sqrt{3\alpha}}  \right)^4    \right) \right ] \nonumber \\
 =&  \frac{1}{2} (\bm{\Phi} + \bar{\bm{\Phi}})^2 + \frac{1 + 8 c_2 + 6 c_2^2 - 12 c_4}{36 \alpha (1 + 2 c_2)} (\bm{\Phi} + \bar{\bm{\Phi}})^4 + \cdots,
 \end{align}
 which is same as eq.~\eqref{K_shift-symmetric} up to higher order terms if the coefficient of the second term is negative (and large). As a side remark, we note that $\bm{T}$ satisfies $\left( \frac{\bm{T}}{\bar{\bm{T}}} -1 \right)^3 = 0$ in the low energy limit as a consequence of $(\bm{\Phi}+\bar{\bm{\Phi}})^3 = 0$.
 
 In summary, the description by the constrained superdield that satisfies $(\bm{\Phi}+ \bar{\bm{\Phi}})^3 = 0$ is valid when $\phi$ is strongly stabilized by the K\"ahler curvature.  The apparent form can be different from eq.~\eqref{K_shift-symmetric} as in eq.~\eqref{modified_hyperbolic_geometry}, but it should be essentially equivalent in the canonical field basis. 

We have just seen an alternative parametrization in terms of $\bm{T}$ rather than $\bm{\Phi}$.  Two other parametrizations have been discussed in Ref.~\cite{Aldabergenov:2021obf}.\footnote{Our notations for superfields are different from Ref.~\cite{Aldabergenov:2021obf}.  Their $\bm{S}$, $\bm{Z}$, $\bm{T}$, and $\bm{X}$ are our $\bm{\Phi}$, $\bm{Z}$, $\bm{A}$, and $\bm{X}$, respectively.  We used $\bm{S}$ for the standard nilpotent superfield and $\bm{T}$ in eq.~\eqref{modified_hyperbolic_geometry}.} In the terminology of Ref.~\cite{Aldabergenov:2021obf}, $\bm{\Phi}$ is the shift-symmetric variable, and $\bm{Z} = \langle Z \rangle e^{\bm{\Phi}}$ (in the natural unit) is the phase-symmetric variable. The latter satisfies $(\bar{\bm{Z}} \bm{Z} - \langle \bar{Z} Z \rangle )^3 = 0$, so the radial mode and the phase mode correspond to the heavy mode and light mode, respectively. 

$\Phi$ is also related to the orthogonal nilpotent superfields $\bm{X}$ and $\bm{A}$ satisfying $\bm{X}^2 = \bm{X} (\bm{A} + \bar{\bm{A}}) = 0 $~\cite{Komargodski:2009rz, Kahn:2015mla, Ferrara:2015tyn, Carrasco:2015iij, DallAgata:2015zxp} via $\bm{\Phi} = \log ( \bm{X} + e^{\bm{A}} )$~\cite{Komargodski:2009rz, Aldabergenov:2021obf}. In Appendix~\ref{sec:comment_gravitino}, we comment on an issue of catastrophic gravitino production~\cite{Hasegawa:2017hgd, Kolb:2021xfn, Kolb:2021nob}, from a perspective we develop in this paper, that arises when one uses the orthogonal nilpotent superfields with a generic superpotential to describe dynamics after inflation. The conclusion of the appendix is that there is no such issue if we start from a UV model like~\eqref{K_shift-symmetric} and reduce it to the low-energy effective theory since the superpotential becomes of a particular form~\cite{Komargodski:2009rz}. The result is consistent with the analysis from another perspective~\cite{Terada:2021rtp}.

\section{Multiple superfields \label{sec:multi}}

In this section, we study a model with multiple superfields and obtain novel constraints on superfields, again without neglecting derivative terms.
It turns out that expressions are much more complicated than those in the single-superfield case in Section~\ref{sec:single}, so we focus on a particular setup with two superfields well-motivated from the perspective of inflation. 

\subsection{Stabilizer model with shift symmetry and quintic constraints \label{sec:stabilizer_model}}
One of the mechanisms of inflation in supergravity utilizes a stabilizer superfield $\bm{S}$ such that $S = 0$ during inflation at least approximately~\cite{Kawasaki:2000yn, Kawasaki:2000ws, Kallosh:2010ug, Kallosh:2010xz, Kallosh:2011qk}.  The superpotential is supposed to be proportional to $\bm{S}$.  This ensures that the negative term in supergravity is no longer harmful.  The superpotential has the following form, 
\begin{align}
    W(\bm{\Phi}, \bm{S}) =& \bm{S} f(\bm{\Phi}), \label{simple_stabilizer_superpotential}
\end{align}
where $f(\bm{\Phi})$ is a holomorphic function of the inflaton superfield $\bm{\Phi}$. One can think of this as the previously constant supersymmetry breaking parameter $f$, which is now slowly changing with $\Phi$. 
In addition, it is typically assumed that the K\"ahler potential has an approximate shift symmetry of the inflaton as in the previous section: $K(\bm{\Phi}, \bm{S}, \bar{\bm{\Phi}}, \bar{\bm{S}}) = K(\bm{\Phi}+\bar{\bm{\Phi}}, \bm{S}, \bar{\bm{S}})$~\cite{Kawasaki:2000yn, Kawasaki:2000ws, Kallosh:2010xz, Kallosh:2011qk}.

As we briefly mentioned before Section~\ref{sec:cubic}, the stabilizer superfield is sometimes replaced by the nilpotent superfield in the literature.  This is usually done by hand, but whether one obtains the nilpotent condition $\bm{S}^2=0$ depends on the details of the UV coupling~\cite{Dudas:2011kt, Antoniadis:2011xi, Antoniadis:2012ck, Ghilencea:2015aph} (but see also Ref.~\cite{Cribiori:2017ngp}).  Moreover, it becomes invalid in the inflaton oscillation regime if the potential $V \simeq |F^S|^2$ vanishes. We address these points simultaneously.

In the case of the single superfield $\bm{\Phi}$, we saw that the denominator has the kinetic energy of the inflaton $\varphi$.  This motivates us to consider a UV theory that lets the inflaton kinetic energy enter the denominator of $S$.  For this to happen, $S$ must feel the supersymmetry breaking by the kinetic energy.  Therefore, we couple $\bm{S}$ and $\bm{\Phi}$ in the K\"ahler potential.  Of course, if we couple both the real and imaginary parts of $\Phi$ to $S$, both are stabilized and inflation does not occur.  This consideration guides us again to the (approximate) shift symmetry of $\bm{\Phi}$. 

Specifically, we consider the following K\"ahler potential,
\begin{align}
K = \frac{1}{2} (\bm{\Phi} + \bar{\bm{\Phi}})^2 + \bar{\bm{S}}\bm{S}  - \frac{1}{4 \Lambda^2_S}(\bar{\bm{S}}\bm{S})^2 - \frac{1}{2 \Lambda^2_{S\phi}} \bar{\bm{S}}\bm{S} (\bm{\Phi} + \bar{\bm{\Phi}})^2 - \frac{1}{4!\Lambda^2_\phi} (\bm{\Phi} + \bar{\bm{\Phi}})^4.
\end{align}
For the inflationary model building with a stabilizer field $S$, we assume an (approximate) $R$-symmetry under which $\bm{S}$ is charged.  This explains the absence of terms like $\bm{S}^2 \bar{\bm{S}}$, $\bm{S}^2 (\bm{\Phi} + \bar{\bm{\Phi}})^2$, etc. 
We also assumed that the vacuum expectation value of the real part $\phi$ is negligible.  For this purpose, we neglected odd terms of $(\bm{\Phi} + \bar{\bm{\Phi}})$ in the K\"ahler potential.  Alternatively, one can say that we imposed the (approximate) $\mathbb{Z}_2$ symmetry.   We could consider a more general K\"ahler potential $K(\bm{\Phi}, \bm{S}, \bar{\bm{\Phi}}, \bar{\bm{S}}) = K(\bm{\Phi}+\bar{\bm{\Phi}}, \bm{S}, \bar{\bm{S}})$, but the above form is simple and well motivated. 

In the limit $1/\Lambda^2_{S\phi} \to 0$, $\bm{S}$ and $\bm{\Phi}$ decouple in global supersymmetry, so $\bm{S}$ reduces to the nilpotent superfield ($\bm{S}^2 = 0$) and $\bm{\Phi}$ satisfies the constraint discussed in Section~\ref{sec:cubic} ($(\bm{\Phi}+\bar{\bm{\Phi}})^3 = 0$). In the presence of nonzero $1/\Lambda^2_{S\phi}$, both constraints are modified as we see below. 

The parts of the equations of motion for $S$ and $\phi$ which are enhanced by the K\"ahler curvature are as follows:
\begin{align}
0 = & - \frac{1}{\Lambda^2_S} S \left(F^S \bar{F}^{\bar{S}} - \frac{i}{2} \left(\chi^S \sigma^\mu \partial_\mu \bar{\chi}^{\bar{S}} + \bar{\chi}^{\bar{S}} \bar{\sigma}^\mu \partial_\mu \chi^S \right)  \right) - \frac{1}{\Lambda^2_{S\phi}} S  \left( F^\Phi \bar{F}^{\bar{\Phi}} - \frac{i}{2} \left(\chi^\Phi \sigma^\mu \partial_\mu \bar{\chi}^{\bar{\Phi}} + \bar{\chi}^{\bar{\Phi}} \bar{\sigma}^\mu \partial_\mu \chi^\Phi \right)  \right) \nonumber \\
& - \frac{\sqrt{2}}{\Lambda^2_{S\phi}} \phi \left( F^S \bar{F}^{\bar{\Phi}} - \frac{i}{2} \left(\chi^S \sigma^\mu \partial_\mu \bar{\chi}^{\bar{\Phi}} + \bar{\chi}^{\bar{\Phi}} \bar{\sigma}^\mu \partial_\mu \chi^S \right)  \right) +  \left(- \frac{1}{\Lambda^2_{S\phi}} \phi^2 - \frac{1}{\Lambda^2_S}\bar{S}S \right) \Box S - \frac{\sqrt{2}}{\Lambda^2_{S\phi}} S \phi \Box \Phi \nonumber \\
&  - \frac{1}{\Lambda^2_S}\bar{S} \partial^\mu S \partial_\mu S - \frac{2\sqrt{2}}{\Lambda^2_{S\phi}} \phi \partial^\mu S \partial_\mu \Phi - \frac{1}{\Lambda^2_{S\phi}} S \partial^\mu \Phi \partial_\mu \Phi \nonumber \\
&- \frac{i}{\Lambda^2_S} \chi^S \sigma^\mu \bar{\chi}^{\bar{S}} \partial_\mu S -\frac{i}{\Lambda^2_{S\phi}} \chi^S\sigma^\mu \bar{\chi}^{\bar{\Phi}} \partial_\mu \Phi - \frac{i}{\Lambda^2_{S\phi}} \chi^\Phi \sigma^\mu \bar{\chi}^{\bar{\Phi}} \partial_\mu S \nonumber \\
& - \frac{i}{2 \Lambda^2_S} S \partial_\mu (\chi^S \sigma^\mu \bar{\chi}^{\bar{S}}) - \frac{i}{\sqrt{2}\Lambda^2_{S\phi}}\phi \partial_\mu (\chi^S \sigma^\mu \bar{\chi}^{\bar{\Phi}}) - \frac{i}{2\Lambda^2_{S\phi}} S \partial_\mu (\chi^\Phi \sigma^\mu \bar{\chi}^{\bar{\Phi}}) \nonumber \\
& + \frac{1}{2\Lambda^2_{S\phi}} F^S \bar{\chi}^{\bar{\Phi}} \bar{\chi}^{\bar{\Phi}} + \frac{1}{2\Lambda^2_S}\bar{F}^{\bar{S}}\chi^S \chi^S + \frac{1}{\Lambda^2_{S\phi}} \bar{F}^{\bar{\Phi}} \chi^\Phi \chi^S, \\
0 =& -\frac{\sqrt{2}}{\Lambda^2_\phi} \phi  \left( F^\Phi \bar{F}^{\bar{\Phi}} - \frac{i}{2} \left(\chi^\Phi \sigma^\mu \partial_\mu \bar{\chi}^{\bar{\Phi}} + \bar{\chi}^{\bar{\Phi}} \bar{\sigma}^\mu \partial_\mu \chi^\Phi \right)  \right)   -\frac{\sqrt{2}}{\Lambda^2_{S\phi}} \phi  \left(F^S \bar{F}^{\bar{S}} - \frac{i}{2} \left(\chi^S \sigma^\mu \partial_\mu \bar{\chi}^{\bar{S}} + \bar{\chi}^{\bar{S}} \bar{\sigma}^\mu \partial_\mu \chi^S \right)  \right) \nonumber  \\
& - \frac{1}{\Lambda^2_{S\phi}} S \left( F^\Phi \bar{F}^{\bar{S}} - \frac{i}{2} \left(\chi^\Phi \sigma^\mu \partial_\mu \bar{\chi}^{\bar{S}} + \bar{\chi}^{\bar{S}} \bar{\sigma}^\mu \partial_\mu \chi^\Phi \right)  \right)  - \frac{1}{\Lambda^2_{S\phi}} \bar{S} \left( F^S \bar{F}^{\bar{\Phi}} - \frac{i}{2} \left(\chi^S \sigma^\mu \partial_\mu \bar{\chi}^{\bar{\Phi}} + \bar{\chi}^{\bar{\Phi}} \bar{\sigma}^\mu \partial_\mu \chi^S \right)  \right) \nonumber \\
& - \frac{1}{\sqrt{2}\Lambda^2_\phi} \phi \left(\partial^\mu \phi \partial_\mu \phi - \partial^\mu \varphi \partial_\mu \varphi \right)  -\frac{1}{\sqrt{2}\Lambda^2_{S\phi}} \partial^\mu \phi \partial_\mu (\bar{S} S) \nonumber \\
& - \frac{i}{\sqrt{2}\Lambda^2_{S\phi}} \partial^\mu \varphi (\bar{S}\partial_\mu S - S \partial_\mu \bar{S}) - \frac{1}{\sqrt{2} \Lambda^2_{S\phi}} \phi \left(\partial^\mu S \partial_\mu S + \partial^\mu \bar{S} \partial_\mu \bar{S} \right)  \nonumber \\
& - \frac{1}{\sqrt{2}} \left( \frac{1}{\Lambda^2_{S\phi}}S\bar{S} + \frac{1}{\Lambda^2_\phi} \phi^2  \right) \Box \phi - \frac{1}{\sqrt{2}\Lambda^2_{S\phi}} \phi (\bar{S} \Box S + S \Box \bar{S} )  \nonumber \\
& + \frac{1}{\sqrt{2}\Lambda^2_\phi} \chi^\Phi \sigma^\mu \bar{\chi}^{\bar{\Phi}} \partial_\mu \varphi + \frac{1}{\sqrt{2}\Lambda^2_{S\phi}} \chi^S \sigma^\mu \bar{\chi}^{\bar{S}} \partial_\mu \varphi  - \frac{i}{2\Lambda^2_{S\phi}} \chi^\Phi \sigma^\mu \bar{\chi}^{\bar{S}} \partial_\mu S + \frac{i}{2\Lambda^2_{S\phi}} \chi^S \sigma^\mu \bar{\chi}^{\bar{\Phi}} \partial_\mu \bar{S}  \nonumber \\
& + \frac{1}{2 \Lambda^2_\phi} \left( F^\Phi \bar{\chi}^{\bar{\Phi}}\bar{\chi}^{\bar{\Phi}} + \bar{F}^{\bar{\Phi}} \chi^\Phi \chi^\Phi \right)+ \frac{1}{\Lambda^2_{S\phi}} \left( F^S \bar{\chi}^{\bar{S}}\bar{\chi}^{\bar{\Phi}} + \bar{F}^{\bar{S}} \chi^S \chi^\Phi \right) .
\end{align}
It is straightforward to solve these equations.  At the quadratic order of fermions, the solution is
\begin{align}
\begin{pmatrix}
S \\
\bar{S} \\
\sqrt{2} \phi
\end{pmatrix}
=& 
\frac{1}{2 m^2_{\bar{S}S} (  m^2_{\bar{S}S} m^2_{\phi\phi} -  |m^2_{S\phi}|^2)} 
\begin{pmatrix}
2 m^2_{\bar{S}S} m^2_{\phi\phi} - |m^2_{S\phi}|^2 & m^4_{\bar{S}\phi} & - 2 m^2_{\bar{S}S} m^2_{\bar{S}\phi} \\
m^4_{S \phi} & 2 m^2_{\bar{S}S} m^2_{\phi\phi} - |m^2_{S\phi}|^2 & - 2 m^2_{\bar{S}S} m^2_{S\phi} \\
- 2 m^2_{\bar{S}S} m^2_{S\phi}  &  - 2 m^2_{\bar{S}S} m^2_{\bar{S}\phi} & 4 m^4_{\bar{S}S} 
\end{pmatrix}
\begin{pmatrix}
A \\
\bar{A} \\
B
\end{pmatrix},\label{Sphi_sol_leading}
\end{align}
where 
\begin{align}
m^2_{\bar{S}S} \equiv &  \frac{1}{\Lambda_S^2} F^S \bar{F}^{\bar{S}} + \frac{1}{\Lambda_{S\phi}^2} \left(- \frac{1}{2}\partial^\mu \varphi \partial_\mu \varphi + F^\Phi \bar{F}^{\bar{\Phi}} \right) , \\
m^2_{\bar{S}\phi}  \equiv  & \frac{2}{\Lambda^2_{S\phi}} F^S \bar{F}^{\bar{\Phi}}, \\
m^2_{S\phi} \equiv & \frac{2}{\Lambda^2_{S\phi}} F^\Phi \bar{F}^{\bar{S}}, \\
m^2_{\phi\phi} \equiv &  \frac{2}{\Lambda_\phi^2} \left(- \frac{1}{2}\partial^\mu \varphi \partial_\mu \varphi + \bar{F}^{\bar{\Phi}} F^\Phi \right) + \frac{2}{\Lambda_{S\phi}^2} \bar{F}^{\bar{S}}F^S, \\
A \equiv & \frac{1}{2 \Lambda_S^2} \bar{F}^{\bar{S}} \chi^S\chi^S + \frac{1}{2\Lambda_{S\phi}^2} F^S \bar{\chi}^{\bar{\Phi}}\bar{\chi}^{\bar{\Phi}} + \frac{1}{\Lambda_{S\phi}^2} \bar{F}^{\bar{\Phi}} \chi^\Phi \chi^S + \frac{1}{\sqrt{2}\Lambda_{S\phi}^2} \partial_\mu \varphi \chi^S \sigma^\mu \bar{\chi}^{\bar{\Phi}}, \\
B \equiv & \frac{1}{2\Lambda_\phi^2} \left(  F^\Phi \bar{\chi}^{\bar{\Phi}}\bar{\chi}^{\bar{\Phi}} + \bar{F}^{\bar{\Phi}} \chi^\Phi \chi^\Phi     \right)  + \frac{1}{\Lambda_{S\phi}^2} \left( F^S \bar{\chi}^{\bar{S}}\bar{\chi}^{\bar{\Phi}} + \bar{F}^{\bar{S}} \chi^S \chi^\Phi \right) \nonumber \\
& + \frac{1}{\sqrt{2}\Lambda_\phi^2} \partial_\mu \varphi \chi^\Phi \sigma^\mu \bar{\chi}^{\bar{\Phi}} + \frac{1}{\sqrt{2}\Lambda_{S\phi}^2} \partial_\mu \varphi \chi^S \sigma^\mu \bar{\chi}^{\bar{S}} .
\end{align}
It should be emphasized that the effective mass parameters $m^2_{\bar{S}S}$ and $m^2_{\phi\phi}$ depend on the kinetic term of the inflaton as well as the $F$-term supersymmetry breaking terms. 

Note that both the expansions of $S$ and $\phi$ start with linear combinations of $\chi^\Phi \chi^\Phi$, $\chi^\Phi \chi^S$, $\chi^S \chi^S$, $\chi^\Phi \sigma^\mu \bar{\chi}^{\bar{\Phi}}$, $\chi^\Phi \sigma^\mu \bar{\chi}^{\bar{S}}$,  $\chi^S \sigma^\mu \bar{\chi}^{\bar{S}}$, and their conjugates generically with nonzero coefficients. 
That is, $S$ and $\phi$ are proportional to at least quadratic fermion terms without derivatives. This fact does not change even if we take into account higher-order fermionic terms.  
These imply that the fifth power of them vanish, so
\begin{align}
\phi^5 = \phi^4 S = \phi^3 S^2 = \phi^3 S \bar{S} = \phi^2 S^3 = \phi^2 S^2 \bar{S} = \phi S^4 = \phi S^3 \bar{S} = \phi S^2 \bar{S}^2 = S^5 = S^4 \bar{S} = S^3 \bar{S}^2 = 0,\label{quintic_comp}
\end{align}
and their complex conjugates are satisfied. Furthermore, the supersymmetry transformation of these constraints are compatible with these constraints.  Because of supersymmetry, the corresponding equations are satisfied by superfields too,
\begin{align}
    &(\bm{\Phi} + \bar{\bm{\Phi}})^5 = (\bm{\Phi} + \bar{\bm{\Phi}})^4 \bm{S} = (\bm{\Phi} + \bar{\bm{\Phi}})^3 \bm{S}^2 = (\bm{\Phi} + \bar{\bm{\Phi}})^3 \bm{S} \bar{\bm{S}} = (\bm{\Phi} + \bar{\bm{\Phi}})^2 \bm{S}^3 = (\bm{\Phi} + \bar{\bm{\Phi}})^2 \bm{S}^2 \bar{\bm{S}} \nonumber \\
    &= (\bm{\Phi} + \bar{\bm{\Phi}}) \bm{S}^4 = (\bm{\Phi} + \bar{\bm{\Phi}}) \bm{S}^3 \bar{\bm{S}} = (\bm{\Phi} + \bar{\bm{\Phi}}) \bm{S}^2 \bar{\bm{S}}^2 = \bm{S}^5 = \bm{S}^4 \bar{\bm{S}} = \bm{S}^3 \bar{\bm{S}}^2 = 0. \label{quintic_constraints}
\end{align}
Some of the constraints further imply a stronger constraint if we assume non-vanishing $F$-term: $D^2 \bm{S} \neq 0$ and/or $D^2 \bm{\Phi} \neq 0$. For example, $\bm{S}^4 \bar{\bm{S}} = 0$ with $\bar{D}^2 \bar{\bm{S}}\neq 0$ implies $\bm{S}^4 = 0$. We have explicitly checked $S^4 = 0$ in terms of the  component fields. 

The same type of the quintic constraint as $(\bm{\Phi}+\bar{\bm{\Phi}})^5 = 0$ on an $\mathcal{N}=2$ superfield were recently proposed in Refs.~\cite{Aldabergenov_Talk, Aldabergenov:2021rxz}.

Similarly to the previous examples, we view these quintic constraints as consequences of the superspace equations of motion 
\begin{align}
    0 = & \mathscr{D}^2 \left[ \frac{1}{\Lambda^2_S} \bar{\bm{S}}\bm{S}^2 + \frac{1}{\Lambda^2_{S\phi}} \bm{S} (\bm{\Phi} + \bar{\bm{\Phi}})^2 \right ], \\
    0 = & (\mathscr{D}^2 + \bar{\mathscr{D}}^2 ) \left [ \frac{1}{\Lambda^2_{S\phi}}\bar{\bm{S}}\bm{S}(\bm{\Phi}+\bar{\bm{\Phi}}) + \frac{1}{6 \Lambda^2_\phi}(\bm{\Phi}+ \bar{\bm{\Phi}})^3 \right].
\end{align}
Applying superderivatives as in the previous examples, we obtain
\begin{align}
& \bm{S} \left( \frac{2}{\Lambda^2_S} (\bar{\mathscr{D}}^2 \bar{\bm{S}} \mathscr{D}^2 \bm{S} + 2 \bar{\mathscr{D}} \bar{\bm{S}} \bar{\mathscr{D}} \mathscr{D}^2 \bm{S} ) + \frac{2}{\Lambda^2_{S\phi}} (2 \bar{\mathscr{D}}\mathscr{D} \bm{\Phi} \bar{\mathscr{D}} \mathscr{D} \bm{\Phi} + \bar{\mathscr{D}}^2 \bar{\bm{\Phi}} \mathscr{D}^2 \bm{\Phi} + 2 \bar{\mathscr{D}} \bar{\bm{\Phi}} \bar{\mathscr{D}} \mathscr{D}^2 \bm{\Phi}  ) \right)  \nonumber \\
& + \frac{2}{\Lambda^2_{S\phi}} (\bm{\Phi} + \bar{\bm{\Phi}}) \left( \bm{S} \bar{\mathscr{D}}^2 \mathscr{D}^2 \bm{\Phi} +4 \bar{\mathscr{D}}\mathscr{D} \bm{S} \bar{\mathscr{D}} \mathscr{D} \bm{\Phi} +2 \bar{\mathscr{D}} \bar{\bm{\Phi}} \bar{\mathscr{D}} \mathscr{D}^2 \bm{S} + \mathscr{D}^2 \bm{S} \bar{\mathscr{D}}^2 \bar{\bm{\Phi}} \right)   \nonumber \\
=& - \frac{2}{\Lambda^2_S} \left( \mathscr{D} \bm{S} \mathscr{D} \bm{S} \bar{\mathscr{D}}^2 \bar{\bm{S}} + 4 \bar{\mathscr{D}}\bar{\bm{S}} \bar{\mathscr{D}} \mathscr{D} \bm{S} \mathscr{D} \bm{S}+ 2 \bar{\bm{S}} \bar{\mathscr{D}} \mathscr{D} \bm{S} \bar{\mathscr{D}} \mathscr{D} \bm{S}  +   \bm{S} \bar{\bm{S}} \bar{\mathscr{D}}^2 \mathscr{D}^2 \bm{S}  \right) \nonumber \\
& - \frac{1}{\Lambda^2_{S\phi}} \left( 4 \mathscr{D} \bm{S} \mathscr{D} \bm{\Phi} \bar{\mathscr{D}}^2 \bar{\bm{\Phi}} +   2 \bar{\mathscr{D}} \bar{\bm{\Phi}} \bar{\mathscr{D}} \bar{\bm{\Phi}} \mathscr{D}^2 \bm{S}  + 8 \bar{\mathscr{D}} \bar{\bm{\Phi}} \mathscr{D} \bm{S} \bar{\mathscr{D}} \mathscr{D} \bm{\Phi}   + 8 \bar{\mathscr{D}}\bar{\bm{\Phi}} \bar{\mathscr{D}} \mathscr{D} \bm{S} \mathscr{D} \bm{\Phi} + (\bm{\Phi} + \bar{\bm{\Phi}})^2 \bar{\mathscr{D}}^2 \mathscr{D}^2 \bm{S} \right), \\
& (\bm{\Phi} + \bar{\bm{\Phi}}) \left( \frac{2}{ \Lambda^2_\phi} (  \bar{\mathscr{D}}^2 \bar{\bm{\Phi}} \mathscr{D}^2 \bm{\Phi} +  (\bar{\mathscr{D}} \mathscr{D} \bm{\Phi} \bar{\mathscr{D}} \mathscr{D} \bm{\Phi} + \text{H.c.}) +  (\bar{\mathscr{D}}\bar{\bm{\Phi}} \bar{\mathscr{D}} \mathscr{D}^2 \bm{\Phi} + \text{H.c.} ))  \right. \nonumber \\
& \qquad  \qquad \left. + \frac{2}{\Lambda^2_{S\phi}} (\bar{\mathscr{D}}^2 \bar{\bm{S}} \mathscr{D}^2 \bm{S} +  ( \bar{\mathscr{D}} \bar{\bm{S}} \bar{\mathscr{D}} \mathscr{D}^2 \bm{S} + \text{H.c.} ) )   \right) \nonumber \\
& + \frac{2 \bm{S}}{\Lambda^2_{S\phi}} \left( \bar{\mathscr{D}}^2 \bar{\bm{S}} \mathscr{D}^2 \bm{\Phi} +  \bar{\mathscr{D}} \bar{\bm{S}} \bar{\mathscr{D}} \mathscr{D}^2 \bm{\Phi} + \mathscr{D}\bm{\Phi} \mathscr{D} \bar{\mathscr{D}}^2 \bar{\bm{S}} + 2 \mathscr{D} \bar{\mathscr{D}} \bar{\bm{S}} \mathscr{D} \bar{\mathscr{D}} \bar{\bm{\Phi}}  \right) 
 + \text{H.c.} \nonumber \\
=& -\frac{1}{2 \Lambda^2_\phi} \left(  4 (\bar{\mathscr{D}} \bar{\bm{\Phi}} \bar{\mathscr{D}} \bar{\bm{\Phi}} \mathscr{D}^2 \bm{\Phi} + \text{H.c.}) + 8 (\bar{\mathscr{D}} \bar{\bm{\Phi}} \bar{\mathscr{D}}\mathscr{D}\bm{\Phi} \mathscr{D} \bm{\Phi} + \text{H.c.})  + (\bm{\Phi} + \bar{\bm{\Phi}})^2 ( \bar{\mathscr{D}}^2 \mathscr{D}^2 \bm{\Phi}  + \text{H.c.}) \right) \nonumber \\
& - \frac{1}{\Lambda^2_{S\phi}} \left( 4 (\bar{\mathscr{D}} \bar{\bm{S}} \bar{\mathscr{D}} \bar{\bm{\Phi}} \mathscr{D}^2 \bm{S} + \text{H.c.})+ 4 (\bar{\mathscr{D}} \bar{\bm{S}}( \bar{\mathscr{D}} \mathscr{D}  \bm{S}  \mathscr{D} \bm{\Phi} +  \bar{\mathscr{D}} \mathscr{D} \bm{\Phi} \mathscr{D} \bm{S} ) + \text{H.c.} ) + \bar{\bm{S}} \bm{S} ( \bar{\mathscr{D}}^2 \mathscr{D}^2 \bm{\Phi} + \text{H.c.}) \right. \nonumber \\
& \qquad  \qquad \left. +   (\bm{\Phi} + \bar{\bm{\Phi}}) (\bar{\bm{S}} \bar{\mathscr{D}}^2 \mathscr{D}^2 \bm{S} + \text{H.c.}) \right).
\end{align}
One can solve these simultaneous equations recursively.

For the inflationary application with the superpotential~\eqref{simple_stabilizer_superpotential}, the purely bosonic part of the $F$-term of $\bm{\Phi}$ ($F^\Phi$) vanishes on-shell.  The equation of motion for $F^\Phi$ is 
\begin{align}
    F^\Phi =  - K^{\bar{S} \Phi} \bar{f}(\bar{\Phi}) - K^{\bar{\Phi} \Phi} \bar{S} f_{\bar{\Phi}}(\bar{\Phi}) + \frac{1}{2} \Gamma^\Phi_{j k} \chi^j \chi^k  . \label{F_Phi_truncated}
\end{align} 
The first and second terms are proportional to $\bar{S} \phi$ and $\bar{S}$, respectively, so there are no purely bosonic terms in $F^\Phi$. 
We see that the leading order solution is 
\begin{align}
 m^2_{\bar{S}S} S &= \frac{1}{2 \Lambda_S^2} \bar{F}^{\bar{S}} \chi^S\chi^S + \frac{1}{2\Lambda_{S\phi}^2} F^S \bar{\chi}^{\bar{\Phi}}\bar{\chi}^{\bar{\Phi}}  + \frac{1}{\sqrt{2}\Lambda_{S\phi}^2} \partial_\mu \varphi \chi^S \sigma^\mu \bar{\chi}^{\bar{\Phi}}, \label{S_sol_truncated_leading} \\
\frac{1}{\sqrt{2}} m^2_{\phi\phi} \phi 
 &= \frac{1}{\Lambda_{S\phi}^2} \left( F^S \bar{\chi}^{\bar{S}}\bar{\chi}^{\bar{\Phi}}  + \bar{F}^{\bar{S}} \chi^S \chi^\Phi \right) + \frac{1}{\sqrt{2}\Lambda_\phi^2} \partial_\mu \varphi \chi^\Phi \sigma^\mu \bar{\chi}^{\bar{\Phi}} + \frac{1}{\sqrt{2}\Lambda_{S\phi}^2} \partial_\mu \varphi \chi^S \sigma^\mu \bar{\chi}^{\bar{S}} . \label{phi_sol_truncated_leading}
\end{align}
In this case, compared to eq.~\eqref{quintic_comp}, lower powers of fields vanish on-shell,
\begin{align}
     \phi^3 S = \phi S^2 = S^3 = 0,
\end{align}
but $\phi^2 S \neq 0$, $\phi^2 S \bar{S} \neq 0$, and $S^2 \bar{S}^2 \neq 0$. We have confirmed this to the full order of fermions by counting the number of $\chi^S$ or $\bar{\chi}^{\bar{\Phi}}$ without derivatives and the number of their conjugates separately. By the same technique, we see that the on-shell  supersymmetry transformation of $S^3$ vanishes to the full order, so $\bm{S}^3 = 0$ is realized in the configuration~\eqref{F_Phi_truncated}. We also observe that the on-shell supersymmetry transformation of $\phi^3 S$ and $\phi S^2$ vanish at the leading order of fermions, which suggests $(\bm{\Phi}+\bar{\bm{\Phi}})^3 \bm{S} = (\bm{\Phi}+\bar{\bm{\Phi}}) \bm{S}^2 = 0$.

\subsection{Implications on cosmological particle production}

Similar to the single-superfield case in eq.~\eqref{phi-only_sol_leading}, our solutions [eq.~\eqref{Sphi_sol_leading} and its on-shell truncation eqs.~\eqref{S_sol_truncated_leading} and \eqref{phi_sol_truncated_leading}] have denominators that depend on the kinetic energy of the inflaton through the dependence on the effective masses $m^2_{\bar{S}S}$ and $m^2_{\phi\phi}$. For simplicity, let us consider the case $\Lambda^2_{S} = \Lambda^2_{\phi} = \Lambda^2_{S\phi} \equiv \Lambda^2$ in eqs.~\eqref{S_sol_truncated_leading} and \eqref{phi_sol_truncated_leading}. In this case, $m^2_{\bar{S}S} = \rho/\Lambda^2 = m^2_{\phi \phi} /2$ where $\rho = - \frac{1}{2} \partial^\mu \varphi \partial_\mu \varphi + |F^S|^2 \simeq \frac{1}{2} \dot{\varphi}^2 + |F^S|^2$ is the energy density (see footnote~\ref{fn:energy_density}), which takes a nonzero value during inflation and the subsequent inflaton oscillation period. The leading terms of the heavy scalar fields become $S = \frac{\bar{F}^{\bar{S}}}{2 \rho} \chi^S \chi^S + \frac{F^S}{2 \rho} \bar{\chi}^{\bar{\Phi}}\bar{\chi}^{\bar{\Phi}} + \frac{\partial_\mu \varphi}{\sqrt{2}\rho} \chi^S \sigma^\mu \bar{\chi}^{\bar{\Phi}}$ 
 and $\phi = \frac{F^S}{\sqrt{2}\rho} \bar{\chi}^{\bar{S}} \bar{\chi}^{\bar{\Phi}} + \frac{\bar{F}^{\bar{S}}}{\sqrt{2} \rho} \chi^S \chi^\Phi + \frac{\partial_\mu \varphi}{2 \rho} (\chi^\Phi \sigma^\mu \bar{\chi}^{\bar{\Phi}} + \chi^S \sigma^\mu \bar{\chi}^{\bar{S}})$.
 It is clear that the denominators are governed by the adiabatic invariant $\rho$. In contrast to the well-studied case with $1/\Lambda^2_{S\phi} \to 0$, where the leading solutions become $S=\frac{1}{2F^S}\chi^S\chi^S$ and $\phi= \frac{1}{-\partial^{\nu}\varphi\partial_{\nu}\varphi}\partial_{\mu}\varphi\chi^{\Phi}\sigma^{\mu}\bar{\chi}^{\bar{\Phi}}$ respectively, the problem of the vanishing denominator does not happen at a finite cosmological time unless the potential minimum is negative, which is anyway problematic~\cite{Linde:2001ae, Felder:2002jk}. Therefore, based on the new set of constraints (eq.~\eqref{Sphi_sol_leading}), we can expand the scope of application of constrained superfields even 
to the situation where either the $F$-term or the kinetic term becomes $0$.
 
 When we remove the assumption of the universal $\Lambda^2$, different linear combinations of the $F$-term and the kinetic term than the energy density appear. However, the qualitative feature is similar. That is, the oscillation behavior of the denominators ($m^2_{\bar{S}S}$ and $m^2_{\phi\phi}$) becomes mild compared to the minimal case as long as the signs of the K\"ahler curvature terms are not changed. 
 
 This mild oscillation behavior may have important implications for particle production during the inflaton coherent oscillation period (or analogous situations for other scalar fields).   In general, non-adiabatic changes of a coupling constant can lead to (resonant) particle production or preheating.  The details may depend on the couplings in the superpotential and canonical normalization etc., but it may well be a sign that violent production would be suppressed in this framework. This will be beneficial in the context of long-lived particles like moduli and gravitinos.
 
 Non-thermal gravitino production during the (p)reheating regime in the presence of the (non-nilpotent) stabilizer field $S$ was discussed in Ref.~\cite{Ema:2016oxl} (see also Refs.~\cite{Maroto:1999ch, Kallosh:1999jj, Giudice:1999yt, Giudice:1999am, Kallosh:2000ve, Nilles:2001ry, Nilles:2001fg} for pioneering work without explicit considerations of the stabilizer field and Ref.~\cite{Roberts:2021plm} for inflatino production during inflation in the presence of the stabilizer field).  One of the key findings there is that gravitino production is suppressed since the gravitino mass approximately vanishes. However, a finite amplitude of the stabilizer field $S$ is gradually generated during the inflaton oscillations, which leads to a finite production rate of gravitinos through the oscillations of the gravitino mass.  In the current setup with approximately equal $\Lambda$'s, the mass term of the stabilizer field $m^2_{\bar{S}S} \simeq \rho / \Lambda^2 = 3 H^2 (M_\text{P}/\Lambda)^2$ is larger than the Hubble scale, so the stabilizer field $S$ is strongly stabilized at the origin even after inflation. This will further suppress the gravitino production rate. 
 
 We also note that the shift-symmetric coupling $(\bm{\Phi} + \bar{\bm{\Phi}})^2 \bar{\bm{S}}\bm{S}$ in the K\"ahler potential does not introduce the gravitino problem due to the coupling between the inflaton and supersymmetry-breaking field in $K$ discussed in Ref.~\cite{Hasegawa:2017nks}.

\section{Generalization \label{sec:generalization}}

So far, we discussed only chiral superfields in global supersymmetry. Some of the discussions are motivated by cosmological applications, which should be ideally described with realistic particle contents and (super)gravitational effects.  We will briefly discuss generalizations including real superfields (gauge-field supermultiplets) and supergravity in this section. 

Before that, we briefly comment on the possibility of other generalizations. In all the examples in this paper, we discuss the decoupling of heavy scalar fields and the effect of the kinetic energy of the light scalar field.  In principle, the same idea will apply to the decoupling of particles with other spins and the kinetic energy of particles with other spins.  For this purpose, one needs superderivative interaction terms. In such a case, one should be careful about the potential emergence of negative-norm states~\cite{DallAgata:2016syy, Fujimori:2016udq}. In a theory with superderivatives, supersymmetry breaking in vacuum by the kinetic energy was recently discussed in Ref.~\cite{Yamada:2021kxv}. Although our analyses in this paper may not directly apply to the cases with superderivatives, it will be interesting to see possible connections. 

It will be also interesting to generalize our strategy to theories with spacetime dimensions other than 4 and/or the number of supersymmetry other than 1 (see footnote~\ref{fn:N=2}).

\subsection{Gauge theory \label{sec:gauge}}
The equation of motion for a complex scalar $X$ in a theory with chiral supermultiplets $(\Phi^i, \chi^i, F^i)$ and gauge-field supermultiplets $(\lambda^A, F^{\mu\nu A}, D^A)$ is
\begin{align}
0 =& K_{\bar{X} i \bar{j}} \left( - \frac{i}{2} \left( \chi^i \sigma^\mu D_\mu \bar{\chi}^{\bar{j}} + \bar{\chi}^{\bar{j}}\bar{\sigma}^\mu D_\mu \chi^i  \right)  + F^i\bar{F}^{\bar{j}} \right) + K_{\bar{X} i} D^\mu D_\mu \Phi^i + K_{\bar{X} ij}  D^\mu \Phi^i  D_\mu \Phi^j \nonumber \\
& + i K_{\bar{X} i j \bar{k}} \chi^j \sigma^\mu \bar{\chi}^{\bar{k}}  D_\mu \Phi^i+ \frac{i}{2} K_{\bar{X}i\bar{j}}  D_\mu (\chi^i \sigma^\mu \bar{\chi}^{\bar{j}} ) 
- \frac{1}{2} K_{\bar{X}i\bar{j}\bar{k}} F^i \bar{\chi}^{\bar{j}}\bar{\chi}^{\bar{k}} - \frac{1}{2} K_{\bar{X}\bar{i}jk} \bar{F}^{\bar{i}} \chi^j \chi^k + \frac{1}{4} K_{\bar{X} ij \bar{k}\bar{\ell}} \chi^i \chi^j \bar{\chi}^{\bar{k}}\bar{\chi}^{\bar{\ell}} \nonumber \\
& + \bar{W}_{\bar{X}\bar{i}}\bar{F}^{\bar{i}}  - \frac{1}{2}\bar{W}_{\bar{X}\bar{i}\bar{j}}\bar{\chi}^{\bar{i}}\bar{\chi}^{\bar{j}} \nonumber \\
& - \frac{1}{8} \bar{h}_{AB \, \bar{X}} \left( F^A_{\mu\nu} F^{\mu\nu \, B} + 2 i \left( \lambda^A \sigma^\mu D_\mu \bar{\lambda}^B + \bar{\lambda}^A \bar{\sigma}^\mu D_\mu \lambda^B \right) - 2 D^A D^B \right) \nonumber \\
& + \frac{i}{8}\bar{h}_{AB \, \bar{X}} \left( F_{\mu\nu}^A \tilde{F}^{\mu\nu \, B} + 2 D_\mu \left( \lambda^A \sigma^\mu \bar{\lambda}^B \right)\right) \nonumber \\
&- D^A \mathcal{P}_{A \, \bar{X}} - \sqrt{2} K_{\bar{X}i\bar{j}} \left( k^{\bar{j}}_A \lambda^A \chi^i + K_A^i \bar{\lambda}^A \bar{\chi}^{\bar{j}} \right) - \sqrt{2} K_{i \bar{j}} (k_A^{\bar{j}})_{\bar{X}} \lambda^A \chi^i \nonumber \\
&-\frac{\sqrt{2}}{4} \bar{h}_{AB \, \bar{X}\bar{i}} \bar{\chi}^{\bar{i}} \bar{\sigma}^{\mu\nu}\bar{\lambda}^A F_{\mu\nu}^B - \frac{\sqrt{2}i}{4} \bar{h}_{AB \, \bar{X}\bar{i}} \bar{\chi}^{\bar{i}}\bar{\lambda}^A D^B - \frac{1}{4} \bar{h}_{AB \, \bar{X}\bar{i}} \bar{F}^{\bar{i}} \bar{\lambda}^A \bar{\lambda}^B + \frac{1}{8} \bar{h}_{AB\, \bar{X}\bar{i}\bar{j}} \bar{\chi}^{\bar{i}} \bar{\chi}^{\bar{j}} \bar{\lambda}^A \bar{\lambda}^B, \label{EoM_gauged}
\end{align}
where $h_{AB}$ is the holomorphic gauge kinetic function, $\mathcal{P}_A$ is the Killing potential, $k_A^i = - i K^{\bar{j}i}\mathcal{P}_{A, \bar{j}}$ is the holomorphic Killing vector of the K\"ahler manifold, and the covariant derivative of the fermion includes the gauge connection, $D_\mu \chi^i = \partial_\mu \chi^i - A_\mu^A \frac{\partial k_A^i}{\partial \Phi^j} \chi^j$ where $A_\mu^A$ is the gauge field, but the K\"ahler connection is stripped away to make the whole expression simpler. 
The upper three lines are just covariantizations of the previous result.
The lower four lines are the new contributions from gauge fields, gauginos, and auxiliary $D$-terms. 

The structure of the fourth line looks similar to that of the first line, but these terms alone cannot stabilize the field $X$ because they give holomorphic mass terms $M_{XX}$ and $M_{\bar{X}\bar{X}}$, that is, tachyonic mass contributions.
Nevertheless, in the presence of both $F$-term and $D$-term supersymmetry breaking, it is straightforward to include the additional terms to obtain new constraints in terms of component fields. However, the $D$-term/gauge-field parts significantly affect the constraints only when the relevant coefficient in the gauge kinetic function is as large as the K\"ahler curvature.\footnote{ \label{fn:N=2}
The K\"ahler potential and the gauge kinetic function are related via the holomorphic prepotential in $\mathcal{N}=2$ supersymmetry. The constraints associated to $\mathcal{N}=2 \to 0$ breaking are higher-order polynomials~\cite{Dudas:2017sbi, Aldabergenov_Talk, Aldabergenov:2021rxz} than in the case of $\mathcal{N}= 1 \to 0$ or $\mathcal{N}=2 \to 1$~\cite{Antoniadis:2017jsk}. The situation is analogous to this case.
} 

As we have repeatedly seen, shift symmetry plays an important role. The equation of motion for the real part of the shift-symmetric superfield $\Phi$ is obtained by simply taking the real part of eq.~\eqref{EoM_gauged} with $\Phi$ replacing $X$.   The holomorphic gauge kinetic function $h_{AB}$ can depend linearly on $\Phi$ without breaking the shift symmetry at the perturbative level.  In such a case, however, $h_{AB \, \Phi} = \text{(real const.)}  $ and $h_{AB \, \Phi i } = 0$, so many terms vanish in the equation of motion:
\begin{align}
0 = & \dots  - \frac{1}{8} \left( \bar{h}_{AB \, \bar{\Phi}} + h_{AB \, \Phi} \right) \left( F^A_{\mu\nu} F^{\mu\nu \, B} + 2 i \left( \lambda^A \sigma^\mu D_\mu \bar{\lambda}^B + \bar{\lambda}^A \bar{\sigma}^\mu D_\mu \lambda^B \right) - 2 D^A D^B \right) \nonumber \\
&- D^A \left( \mathcal{P}_{A \, \bar{\Phi}} + \mathcal{P}_{A \, \Phi} \right)  - \sqrt{2} \left( K_{\Phi i \bar{j}} +  K_{\bar{\Phi}i\bar{j}} \right) \left( k^{\bar{j}}_A \lambda^A \chi^i + K_A^i \bar{\lambda}^A \bar{\chi}^{\bar{j}} \right)  - \sqrt{2} K_{i \bar{j}} \left( (k_A^{\bar{j}})_{\bar{\Phi}} \lambda^A \chi^i + (k_A^i)_\Phi \bar{\lambda}^A \bar{\chi}^{\bar{j}} \right) ,
\end{align}
where dots denote the non-gauge ($F$-term) part.
The bosonic terms in the first line shift the vacuum expectation value of the field $\phi$, which can be subtracted by field redefinition. On the other hand, the fermionic term (gaugino bilinear) in the first line affects the solution at the quadratic level.  This means, strictly speaking, that the constraint becomes a higher-order polynomial equation.  Again, this happens practically when the coefficient of the gauge kinetic function is comparable to the K\"ahler curvature. Otherwise, the effects of these new terms appear as small corrections to the original constraint suppressed by the K\"ahler curvature.  Other terms affect the solution at higher orders.

\subsection{Supergravity \label{sec:sugra}}
If we consider large-field inflation, it is important to consider the effects of coupling to (super)gravity. Constraint superfields in supergravity have been discussed in early days~\cite{Lindstrom:1979kq} and recently, e.g., in Refs.~\cite{Farakos:2013ih, Dudas:2015eha, DallAgata:2015zxp, Kallosh:2015pho, Bandos:2016xyu, Cribiori:2016qif, Cribiori:2017laj}.  Here, we begin with the standard UV setup with a linearly realized supergravity rather than imposing a constraint by hand.  We take the supergravity supermultiplet on-shell while the matter supermultiplets off-shell (see Ref.~\cite{Freedman:2017obq} for off-shell supergravity).  For simplicity, we only consider chiral superfields for matter.  

The equation of motion for a complex scalar field $X$ in supergravity is given in eq.~\eqref{EoM_X_SUGRA} in Appendix~\ref{sec:SUGRA_EoM}.   
New terms proportional to the K\"ahler curvature appear, but these are suppressed by (the ratio between the additional fields and) the reduced Planck scale.  Though there are various contributions at the quartic order in fermions (including gravitino), we do not find any new contributions at the quadratic level. Thus, we do not find any significant differences in the supergravity extension.

\section{Conclusion \label{sec:conclusion}}

In this paper, we have studied constrained superfields in dynamical/cosmological backgrounds where derivatives of scalar fields such as the kinetic energy are sizable.  We have started from the models with supersymmetry realized linearly.  The K\"ahler potential has an enhanced curvature with a shift symmetry, which makes some scalar fields heavy enough to decouple but keeps other fields light.  We have integrated out the heavy modes, i.e., we have solved the equations of motion for the heavy scalars to obtain supersymmetric constraints. The shift-symmetric cubic constraint~\eqref{cubic_constraint} was discussed in the literature, but we have derived it from a UV model explicitly.  We have also derived novel quintic constraints~\eqref{quintic_constraints} for the double superfield case that is motivated in the context of inflation with the stabilizer field.  Though the details depend on how many superfields are coupled by the quartic terms in the K\"ahler potential, the essential thing is that the kinetic energy of light fields as well as the potential energy contributes to the effective masses of the heavy fields. Depending on the sign of the K\"ahler curvature, the effect of the kinetic energy can suppress the non-adiabatic changes of the effective masses and coupling constants in the low-energy theory in which supersymmetry is nonlinearly realized.  We have discussed their cosmological implications. In particular, it is remarkable that the constrained superfields we obtained in this work allow us to describe the low energy effective field theory even during the inflaton oscillation era without violating its validity. This is in sharp contrast to the standard inflation scenarios based on the nilpotent stabilizer superfield.  We have also delineated the generalizations into supersymmetric gauge theories and supergravity.

Key ingredients in our discussion are the shift-symmetric couplings in the K\"ahler curvature terms.  In fact, whether we use constrained superfields or unconstrained superfields is a matter of choice.  In the first place, the shift-symmetric couplings in the K\"ahler potential themselves are optional unless there is tachyonic instability. Nevertheless, if there are indeed the shift-symmetric quartic couplings, it is not only that the description in terms of constrained superfields can be extended to the large kinetic-energy regime but also that there are cosmological advantages as discussed in this paper.  It is also interesting that the shift symmetry is favorable in the context of inflation.  Meanwhile, it is desirable to explain the origin of the shift-symmetric quartic terms by an underlying field-theoretic or stringy mechanism. 

Our strategy can be straightforwardly applied also to cases with multiple superfields breaking supersymmetry. Although calculations to derive explicit forms of the solutions would become more tedious, it is easy to expect the orders of the algebraic constraints by counting the number of fermion species involved. 
When $N$ chiral superfields couple with each other without shift symmetry, the constraints are expected to be $(N+1)$-th order  (see  Refs.~\cite{Dudas:2011kt, Antoniadis:2011xi, Antoniadis:2012ck} for a more precise discussion for small values of $N$).   If all the $N$ superfields respect their shift symmetries, the constraints are expected to be $(2N+1)$-th order. When fields with and without shift symmetries couple each other with generic coefficients, the constraints may depend on how dense the coupling structure is. Our example in Section~\ref{sec:multi} suggests the power of the constraint is still $2N+1$ for a sufficiently large number of shift symmetries. To check these expectations is one of future work.

Most of our cosmological discussions have been motivated by inflation, but constrained superfields in the dynamical background can apply to a broader range of cosmological scenarios and mechanisms including kination, the curvaton mechanism, the kinetic as well as standard axion misalignment mechanisms, the relaxion mechanism, and so on. Constrained superfields in a dynamical/cosmological background are useful tools to describe our dynamically evolving universe. 

\section*{Acknowledgment}
TT thanks Emilian Dudas, Fernando Marchesano, and Luca Martucci for discussions at ``Dark World to Swampland 2021'', The 6th IBS-IFT-MultiDark Workshop. 
This work was supported in part by IBS under the project code, IBS-R018-D1, and in part by Basic Science Research Program through the National Research Foundation of Korea (NRF) funded by the Ministry of Education, Science and
Technology (NRF-2019R1A2C2003738).


\appendix

\section{Equation of motion for generic K\"ahler potential with single superfield \label{sec:general_EOM}}
Here we show the equation of motion of $\phi$ (real part of the scalar component of $\bm{\Phi}$) for a generic K\"ahler potential $K(\Phi,\bar{\Phi})$ and superpotential $W(\Phi)$, which are not necessarily shift symmetric. It is given by
\begin{align}
\nonumber 0= & \operatorname{Re}(K^{(2,1)})\left[\frac{1}{2}\left(\partial_{\mu}\phi\right)^{2}-\frac{1}{2}\left(\partial_{\mu} \varphi\right)^{2}+F \bar{F}-\frac{i}{2} \chi \sigma^{\mu} \partial_{\mu} \bar{\chi}+\frac{i}{2} \partial_{\mu}\chi \sigma^{\mu} \bar{\chi}\right]\\
\nonumber &+\operatorname{Im}(K^{(2,1)})\left[-\partial_{\mu} \phi \partial^{\mu} \varphi+\frac{1}{2}\partial_{\mu}(\chi \sigma^{\mu} \bar{\chi})\right]+\frac{1}{\sqrt{2}}K^{(1,1)}\square\phi-\frac{1}{\sqrt{2}}K^{(2,2)}\partial_{\mu} \varphi \chi \sigma^{\mu} \bar{\chi}\\
&-\frac{1}{2}\operatorname{Re}\left(FK^{(1,3)}\bar{\chi}\bar{\chi}+\bar{F}K^{(2,2)}\chi\chi\right)+\frac{1}{4}\operatorname{Re}(K^{(2,3)})\chi\chi\bar{\chi}\bar{\chi}+\operatorname{Re}\left(FW^{(2)}\right)-\frac{1}{2}\operatorname{Re}\left(W^{(3)} \chi \chi\right),
\end{align}
where we introduced a notation, $K^{(1,2)}\equiv \partial^3 K/\partial\Phi\partial\bar{\Phi}^2$, $W^{(2)}\equiv \partial^2W/\partial \Phi^2$, and so on.
Expanding the equation around a vacuum expectation value as $\Phi=\left\langle \Phi\right\rangle+\tilde{\Phi}$ with $\tilde{\Phi}=\frac{1}{\sqrt{2}}(\tilde{\phi}+i\tilde{\varphi})$, up to the cubic order in $\tilde{\phi}$, we obtain
\begin{align}
\sum_{i\leq j}{\rm{Re}}\left\langle K^{(i,j)}\right\rangle A^{(i,j)}+\sum_{i<j}{\rm{Im}}\left\langle K^{(i,j)}\right\rangle B^{(i,j)}+\left(\sum_{i}\left\langle W^{(i)}\right\rangle C^{(i)}+{\rm{H.c.}}\right)+\mathcal{O}(\tilde{\phi}^3)=0, \label{EOM_phi}  
\end{align}
where 
\begin{align}
A^{(1,1)}=&\frac{1}{\sqrt{2}}\square \tilde{\phi},\\
A^{(1,2)}=&\frac{1}{2}(\partial_{\mu} \tilde{\phi}^{2})-\frac{1}{2}\left(\partial_{\mu} \tilde{\varphi}\right)^{2}+F \bar{F}-\frac{i}{2} \chi \sigma^{\mu} \partial_{\mu} \bar{\chi}+\frac{i}{2} \partial_{\mu}\chi \sigma^{\mu} \bar{\chi} +\tilde{\phi}\square \tilde{\phi}, \\
\nonumber A^{(2,2)}=&\frac{1}{\sqrt{2}}\tilde{\phi}\left[\frac{1}{2}(\partial_{\mu} \tilde{\phi}^{2})-\frac{1}{2}\left(\partial_{\mu} \tilde{\varphi}\right)^{2}+F \bar{F}-\frac{i}{2} \chi \sigma^{\mu} \partial_{\mu} \bar{\chi}+\frac{i}{2} \partial_{\mu}\chi \sigma^{\mu} \bar{\chi}\right]+\frac{1}{2\sqrt{2}}\tilde{\phi}^2\square \tilde{\phi}-\frac{1}{\sqrt{2}}\chi \sigma^{\mu} \bar{\chi} \partial_{\mu} \tilde{\varphi}\\
&-\frac{1}{2}{\rm{Re}}(F\bar{\chi}\bar{\chi}),\\
A^{(1,3)}=&\frac{1}{\sqrt{2}}\tilde{\phi}\left[\frac{1}{2}(\partial_{\mu} \tilde{\phi}^{2})-\frac{1}{2}\left(\partial_{\mu} \tilde{\varphi}\right)^{2}+F \bar{F}-\frac{i}{2} \chi \sigma^{\mu} \partial_{\mu} \bar{\chi}+\frac{i}{2} \partial_{\mu}\chi \sigma^{\mu} \bar{\chi}\right]+\frac{1}{2\sqrt{2}}\tilde{\phi}^2\square \tilde{\phi}-\frac{1}{2}{\rm{Re}}(F\bar{\chi}\bar{\chi}),\\
\nonumber A^{(2,3)}=&\frac{3}{4}\tilde{\phi}^2\left[\frac{1}{2}(\partial_{\mu} \tilde{\phi}^{2})-\frac{1}{2}\left(\partial_{\mu} \tilde{\varphi}\right)^{2}+F \bar{F}-\frac{i}{2} \chi \sigma^{\mu} \partial_{\mu} \bar{\chi}+\frac{i}{2} \partial_{\mu}\chi \sigma^{\mu} \bar{\chi}\right]-\tilde{\phi}\partial_{\mu}\tilde{\varphi}\chi\sigma^{\mu}\bar{\chi}\\
&-\frac{3}{2\sqrt{2}}\tilde{\phi}{\rm{Re}}(F\bar{\chi}\bar{\chi})+\frac{1}{4}\chi\chi\bar{\chi}\bar{\chi},\\
A^{(1,4)}=&\frac{1}{4}\tilde{\phi}^2\left[\frac{1}{2}(\partial_{\mu} \tilde{\phi}^{2})-\frac{1}{2}\left(\partial_{\mu} \tilde{\varphi}\right)^{2}+F \bar{F}-\frac{i}{2} \chi \sigma^{\mu} \partial_{\mu} \bar{\chi}+\frac{i}{2} \partial_{\mu}\chi \sigma^{\mu} \bar{\chi}\right]-\frac{1}{2\sqrt{2}}\tilde{\phi}{\rm{Re}}(F\bar{\chi}\bar{\chi}),\\
A^{(3,3)}=&-\frac{1}{2\sqrt{2}}\tilde{\phi}^2\partial_{\mu}\tilde{\varphi}\chi\sigma^{\mu}\bar{\chi}-\frac{3}{8}\tilde{\phi}^2{\rm{Re}}(F\bar{\chi}\bar{\chi})+\frac{1}{4\sqrt{2}}\tilde{\phi}\chi\chi\bar{\chi}\bar{\chi},\\
A^{(2,4)}=&-\frac{1}{2\sqrt{2}}\tilde{\phi}^2\partial_{\mu}\tilde{\varphi}\chi\sigma^{\mu}\bar{\chi}-\frac{1}{2}\tilde{\phi}^2{\rm{Re}}(F\bar{\chi}\bar{\chi})+\frac{1}{4\sqrt{2}}\tilde{\phi}\chi\chi\bar{\chi}\bar{\chi},\\
A^{(1,5)}=&-\frac{1}{8}\tilde{\phi}^2{\rm{Re}}(F\bar{\chi}\bar{\chi}),\\
A^{(3,4)}=&\frac{3}{16}\tilde{\phi}^2\chi\chi\bar{\chi}\bar{\chi},\\
A^{(2,5)}=&\frac{1}{16}\tilde{\phi}^2\chi\chi\bar{\chi}\bar{\chi},
\end{align}
\begin{align}
&B^{(1,2)}=  \partial_{\mu} \tilde{\phi} \partial^{\mu} \tilde{\varphi}-\frac{1}{2}\partial_{\mu}(\chi \sigma^{\mu} \bar{\chi}),\\
&B^{(1,3)}= \frac{1}{\sqrt{2}}\tilde{\phi}\left[\partial_{\mu} \tilde{\phi} \partial^{\mu} \tilde{\varphi}-\frac{1}{2}\partial_{\mu}(\chi \sigma^{\mu} \bar{\chi})\right]+\frac{1}{2}{\rm{Im}}(F\bar{\chi}\bar{\chi}),\\
&B^{(1,4)}=B^{(2,3)}=\frac{1}{4}\tilde{\phi}^2\left[\partial_{\mu} \tilde{\phi} \partial^{\mu} \tilde{\varphi}-\frac{1}{2}\partial_{\mu}(\chi \sigma^{\mu} \bar{\chi})\right]+\frac{1}{2\sqrt{2}}\tilde{\phi}{\rm{Im}}(F\bar{\chi}\bar{\chi}),\\
&B^{(2,4)}=\frac{1}{4}\tilde{\phi}^2{\rm{Im}}(F\bar{\chi}\bar{\chi}),\\
&B^{(1,5)}=\frac{1}{8}\tilde{\phi}^2{\rm{Im}}(F\bar{\chi}\bar{\chi}),
\end{align}
and 
\begin{align}
&C^{(2)}=\frac{F}{2},\\
&C^{(3)}=\frac{F}{2\sqrt{2}}\tilde{\phi}-\frac{1}{4}\chi\chi,\\
&C^{(4)}=\frac{F}{8}\tilde{\phi}^2-\frac{1}{4\sqrt{2}}\chi\chi\tilde{\phi},\\
&C^{(5)}=-\frac{1}{16}\chi\chi\tilde{\phi}^2,
\label{Wexp}
\end{align}
Note that when $K$ has an exact shift symmetry, all coefficients of $B$-terms vanish, ${\rm{Im}}\left\langle K^{(i,j)}\right\rangle=0$, and there is no distinction between $\Phi$ and $\bar{\Phi}$ superscripts on ${\rm{Re}}\left\langle K^{(i,j)}\right\rangle$, e.g., ${\rm{Re}}\left\langle K^{(2,2)}\right\rangle={\rm{Re}}\left\langle K^{(1,3)}\right\rangle$. Since the K\"ahler curvature is determined by the fourth derivative of the K\"ahler potential, terms from $A^{(2,2)}$ and $A^{(1,3)}$ would be dominant in the large curvature limit in shift symmetric case, which leads to eq.~$\eqref{EoM_phi-only}$. For general case with non-shift symmetric K\"ahler potential, however, there are several corrections with ${\rm{Im}}\left\langle K^{(i,j)}\right\rangle$ ($B$-terms) and superpotential ($C$-terms). As long as they can be treated as small corrections compared to the K\"ahler curvature, the discussion in the main text can be applied.

\section{Comment on the slow gravitino issue \label{sec:comment_gravitino}}

In this appendix, we discuss an issue related to the orthogonal nilpotent superfields $X$ and $A$, which satisfy $\bm{X}^2 = \bm{X}(\bm{A} + \bar{\bm{A}}) = 0$~\cite{Komargodski:2009rz, Kahn:2015mla, Ferrara:2015tyn, Carrasco:2015iij, DallAgata:2015zxp} and related to $\bm{\Phi}$ via  $\bm{\Phi} = \log ( \bm{X} + e^{\bm{A}} )$~\cite{Komargodski:2009rz, Aldabergenov:2021obf}.  An application of the orthogonal nilpotent superfields with a generic superpotential to the dynamics after inflation is known to produce either a breakdown of the effective theory~\cite{Hasegawa:2017hgd} or a catastrophic production of gravitinos due to a significant change of the propagation speed of gravitino~\cite{Hasegawa:2017hgd, Kolb:2021xfn, Kolb:2021nob} (see also Refs.~\cite{Dalianis:2017okk, Dudas:2021njv, Terada:2021rtp, Antoniadis:2021jtg}). Since we have started from a perfectly fine theory~\eqref{K_shift-symmetric} up to the scale around $\Lambda$, it is instructive to see what happens for gravitino in our setup. (Similar or related comments have been given in Refs.~\cite{Hasegawa:2017hgd, Terada:2021rtp}.)  

In the canonical basis, the K\"ahler potential of the constrained superfields is
\begin{align}
    K =& \frac{1}{2} (\bm{\Phi} + \bar{\bm{\Phi}})^2 \nonumber \\
    =& \frac{1}{2}(\bm{A} + \bar{\bm{A}})^2 + \bar{\bm{X}}\bm{X}.
\end{align}
Thus, the minimal shift-symmetric $K$ for $\bm{\Phi}$ simply corresponds to the minimal $K$ for $\bm{A}$ and $\bm{X}$. 
The general superpotential is expressed as 
\begin{align}
    W(\bm{\Phi}) = W(\bm{A}) + \bm{X} W'(\bm{A}) e^{- \bm{A}},
\end{align}
where $W'$ is the derivative of $W$ with respect to its argument. This has a very specific form as a holomorphic function $W(\bm{A}, \bm{X})$~\cite{Komargodski:2009rz}.
This specific structure ensures $|\dot{m}_{3/2}|^2 = 2 |F^X|^2 |\dot{A}|^2$, where $m_{3/2}$ is the gravitino mass, $F^X$ is the $F$-term of $\bm{X}$, and we have used $\langle A + \bar{A} \rangle = \langle X \rangle = \langle e^K \rangle = 0$.\footnote{
As we see in Section.~\ref{sec:sugra} and Appendix~\ref{sec:SUGRA_EoM}, the extension to supergravity does not introduce a big change to the results obtained in global supersymmetry.
} Therefore, the propagation speed of the longitudinal mode of gravitino is same as the speed of light,
\begin{align}
    c_{3/2}^2 = & \frac{\left(|\dot{A}|^2 - |F^X|^2 \right)^2 + | \dot{m}_{3/2}|^2}{\left(|\dot{A}|^2 + |F^X|^2 \right)^2} = 1.
\end{align}
This means that the gravitino issue related to the significant change of the propagation speed is absent in our setup, which is consistent with the analysis in terms of $\bm{\Phi}$~\cite{Terada:2021rtp}. 

\section{Equation of motion in supergravity \label{sec:SUGRA_EoM}}

The equation of motion for the complex scalar field $X$ (without gauge fields) in supergravity is as follows:
\begin{align}
0=&K_{\bar{X} i \bar{j}} \left( - \frac{i}{2} \left( \chi^i \sigma^\mu (\partial_\mu - i A_\mu) \bar{\chi}^{\bar{j}} + \bar{\chi}^{\bar{j}}\bar{\sigma}^\mu (\partial_\mu + i A_\mu) \chi^i  \right)  + F^i\bar{F}^{\bar{j}} \right) + K_{\bar{X} i} \Box \Phi^i + K_{\bar{X} ij} \partial^\mu \Phi^i \partial_\mu \Phi^j \nonumber \\
& + i K_{\bar{X} i j \bar{k}} \chi^j \sigma^\mu \bar{\chi}^{\bar{k}} \partial_\mu \Phi^i+ \frac{i}{2} K_{\bar{X}i\bar{j}} \partial_\mu (\chi^i \sigma^\mu \bar{\chi}^{\bar{j}} ) 
- \frac{1}{2} K_{\bar{X}i\bar{j}\bar{k}} F^i \bar{\chi}^{\bar{j}}\bar{\chi}^{\bar{k}} - \frac{1}{2} K_{\bar{X}\bar{i}jk} \bar{F}^{\bar{i}} \chi^j \chi^k + \frac{1}{4} K_{\bar{X} ij \bar{k}\bar{\ell}} \chi^i \chi^j \bar{\chi}^{\bar{k}}\bar{\chi}^{\bar{\ell}} \nonumber \\
& +\left( e^{K/2} \bar{D}_{\bar{i}} \bar{W} \right)_{\bar{X}} \bar{F}^{\bar{i}}  - \frac{1}{2}  m_{\bar{i}\bar{j}\bar{X}}  \bar{\chi}^{\bar{i}}\bar{\chi}^{\bar{j}} \nonumber \\
& + 3 e^K W \bar{D}_{\bar{X}}\bar{W} - \frac{1}{2} m_{i j \bar{X}} \chi^i \chi^j + \left( e^{K/2} D_i W \right)_{\bar{X}} F^i - \frac{1}{4M_\text{P}^2} K_{\bar{X}i\bar{k}} K_{j \bar{\ell}} \chi^i \chi^j \bar{\chi}^{\bar{k}} \bar{\chi}^{\bar{\ell}} \nonumber \\
& - \frac{K_{i\bar{j}}}{2 M_\text{P}^2} \chi^i \sigma^\mu \bar{\chi}^{\bar{j}} K_{\bar{X}i} \partial_\mu \Phi^i  - \frac{1}{4M_\text{P}^2}  K_{i \bar{j}} K_{\bar{X}} \partial_\mu \left( \chi^i \sigma^\mu \bar{\chi}^{\bar{j}} \right) - \frac{1}{4 M_\text{P}^2} \left( K_{i \bar{j} k} \partial_\mu \Phi^k + K_{i \bar{j} \bar{k}} \partial_\mu \bar{\Phi}^{\bar{k}} \right) K_{\bar{x}} \chi^i \sigma^\mu \bar{\chi}^{\bar{j}} \nonumber \\
& + \frac{i}{4M_\text{P}^2}   K_{\bar{X}i} \partial_\nu \Phi^i  \left( \bar{\psi}_\mu  \bar{\sigma}^{ [ \mu} \sigma^\nu \bar{\sigma}^{\rho ] } \psi_\rho -  \psi_\mu \sigma^{ [ \mu }\bar{\sigma}^\nu \sigma^{\rho ] } \bar{\psi}_\rho \right)  + \frac{i}{8 M_\text{P}^2} K_{\bar{X}} \partial^\mu \left( \bar{\psi}_\mu  \bar{\sigma}^{ [ \mu} \sigma^\nu \bar{\sigma}^{\rho ] } \psi_\rho -  \psi_\mu \sigma^{ [ \mu }\bar{\sigma}^\nu \sigma^{\rho ] } \bar{\psi}_\rho \right)   \nonumber \\
& - \frac{1}{2 M_\text{P}^2} K_{\bar{X}} m_{3/2} \psi_\mu \sigma^{\mu\nu} \psi_\nu - e^{K/2} \left( \bar{W}_{\bar{X}} + \frac{1}{2} K_{\bar{X}} \bar{W} \right) \bar{\psi}_\mu \bar{\sigma}^{\mu\nu} \bar{\psi}_\nu  \nonumber \\
& + \frac{2\sqrt{2}}{\Mpl}
	\left( - K_{\bar{X} i j } \psi_\mu \sigma^{\mu \nu}  \chi^{i} \partial_{\nu}\Phi^j  + K_{\bar{X}i \bar{j}} \bar{\psi}_\mu \bar{\sigma}^{\mu \nu} \bar{\chi}^{\bar{j}}\partial_\nu \Phi^{i}\right) - \frac{2 \sqrt{2}} {M_\text{P}^2}  K_{i \bar{X}}  \partial_\nu \left( \psi_\mu \sigma^{\mu\nu} \chi^i  \right) \nonumber \\
& +\frac{i}{\sqrt{2}M_\text{P}} \bar{\psi}_\mu \bar{\sigma}^\mu \left( \left( e^{K/2} D_i W \right)_{\bar{X}} \chi^i + i K_{\bar{X} i \bar{j}} \sigma^\nu \partial_\nu \Phi^i \bar{\chi}^{\bar{j}} \right) \nonumber \\
&+\frac{i}{\sqrt{2}M_\text{P}} \psi_\mu \sigma^\mu \left( \left( e^{K/2} \bar{D}_{\bar{j}} \bar{W} \right)_{\bar{X}} \bar{\chi}^{\bar{j}} - i K_{\bar{X} i j} \bar{\sigma}^\nu \partial_\nu \Phi^j \chi^i \right)  + \frac{1}{\sqrt{2} M_\text{P}} K_{i\bar{X}} \partial_\nu \left(  \psi_\mu \sigma^\mu  \bar{\sigma}^\nu \chi^i \right) \nonumber \\
& + K_{\bar{X} i\bar{j}}  \left( - \frac{i}{16}e^{-1} \epsilon^{\mu\nu\rho\sigma}  \left( \psi_\mu \sigma_\nu \bar{\psi}_\rho + \bar{\psi}_\mu \bar{\sigma}_\nu \psi_\rho \right)  \bar{\chi}^{\bar{j}} \bar{\sigma}_\sigma \chi^i - \frac{1}{2}  \bar{\psi}_\mu \bar{\chi}^{\bar{j}} \psi^\mu \chi^i   \right),\label{EoM_X_SUGRA}
\end{align}
where $\psi_\mu$ is the gravitino field, $m_{3/2} \equiv e^{K/2} W$ is its mass parameter, $A_\mu \equiv  \frac{i}{4 M_\text{P}^2} \left( K_i \partial_\mu \Phi^i - K_{\bar{i}} \partial_\mu \bar{\Phi}^{\bar{i}}  \right) $ is the on-shell vector  auxiliary field in supergravity, the fermion mass matrix is given by $m_{ij} = e^{K/2} (W_{ij} + K_i W_j + K_j W_i + (K_{ij}+K_i K_j)W - g^{\bar{\ell}k}g_{i j \bar{\ell}} (W_k + K_k W))$, and $ \sigma^{[ \mu} \bar{\sigma}^{\nu} \sigma^{\rho ] }  = \frac{1}{3!} \left( \sigma^{\mu} \bar{\sigma}^{\nu} \sigma^{\rho } + \cdots \right) $ is totally anti-symmetric with respect to indices within the parenthesis.
Some formulas are given below
\begin{align}
\left( e^{K/2} D_i W \right)_{\bar{X}} =& e^{K/2} \left( K_{i \bar{X}} W + \frac{1}{2} K_{\bar{X}} (W_i + K_i W) \right) , \\
\left( e^{K/2} \bar{D}_{\bar{i}} \bar{W} \right)_{\bar{X}} =& e^{K/2} \left( \bar{W}_{\bar{i}\bar{X}} + K_{\bar{i}\bar{X}}\bar{W} + K_{\bar{i}} \bar{W}_{\bar{X}} + \frac{1}{2} K_{\bar{X}} ( \bar{W}_{\bar{i}} + K_{\bar{i}} \bar{W})   \right)  , \\
m_{ij \bar{X}} =& e^{K/2} \left( K_{\bar{X}i}W_j + K_{\bar{X}j}W_i +(K_{\bar{X}ij} + K_{\bar{X}i}K_j + K_i K_{\bar{X}j} ) W \right. \nonumber \\
& \left.  + \frac{1}{2} K_{\bar{X}} \left( W_{ij} + K_i W_j + K_j W_i + (K_{ij} + K_i K_j) W \right) \right) , \\
m_{\bar{i}\bar{j}\bar{X}} = & e^{K/2} \left( \bar{W}_{\bar{X} \bar{i}\bar{j}} +K_{\bar{X} \bar{i}} \bar{W}_{\bar{j}} + K_{\bar{X} \bar{j}} \bar{W}_{\bar{i}} + K_{\bar{i}} \bar{W}_{\bar{X}\bar{j}} + K_{\bar{j}} \bar{W}_{\bar{X} \bar{i}} + (K_{\bar{X}\bar{i}\bar{j}} + K_{\bar{X} \bar{i}}K_{\bar{j}} + K_{\bar{i}} K_{\bar{X}\bar{j}} ) \bar{W}  \right. \nonumber \\
& \left.+ (K_{\bar{i}\bar{j}} + K_{\bar{i}}K_{\bar{j}}) \bar{W}_{\bar{X}}  + \frac{1}{2} K_{\bar{X}} \left( \bar{W}_{\bar{i}\bar{j}} + K_{\bar{i}} \bar{W}_{\bar{j}} + K_{\bar{j}} \bar{W}_{\bar{i}} + (K_{\bar{i}\bar{j}} + K_{\bar{i}}K_{\bar{j}}) \bar{W} \right)  \right).
\end{align}

The first three lines of eq.~\eqref{EoM_X_SUGRA} are a minor modification of the result in global supersymmetry: the fermion kinetic term includes the auxiliary vector $A_\mu$, and the derivatives of the superpotential are associated with the factor $e^{K/2}$. 
From the fourth line, the terms are qualitatively new contributions in supergravity.  

\bibliographystyle{utphys}
\bibliography{ref.bib}

\end{document}